\documentclass[twocolumn,pre,amssymb,floatfix]{revtex4}

\usepackage{times}
\usepackage{subfigure}
\usepackage[sort&compress]{natbib}
\usepackage{epsfig}
\usepackage{graphicx}

\usepackage{fancyhdr}
\usepackage{latexsym}
\usepackage{graphicx}
\usepackage{amsmath,amsmath,amsfonts}

\def\ee{\mathrm{e}}

\def\rv{{\bf r}}
\def\kv{{\bf k}}
\def\qv{{\bf q}}

\def\Rv{{\bf R}}

\def\Iv{{\bf I}}

\def\Uh{U}
\def\Uhb{{\bf \Uh}}

\def\Ss{\Omega}
\def\Ssh{\Ss}

\def\Sshb{{\bf \Ss}}

\def\Debye{g}

\graphicspath{{figs/}}

\begin{document}
\title{Renormalized One-loop Theory of Correlations in Polymer Blends}
\author{Jian Qin and David C. Morse}
\affiliation{
Department of Chemical Engineering and Materials Science, 
University of Minnesota, 
421 Washington Ave. S.E., Minneapolis, MN 55455 }
\date{\today}
\begin{abstract} 
The renormalized one-loop theory is a coarse-grained theory of corrections 
to the self-consistent field theory (SCFT) of polymer liquids, and to the 
random phase approximation (RPA) theory of composition fluctuations. We 
present predictions of corrections to the RPA for the structure function 
$S(k)$ and to the random walk model of single-chain statics in binary 
homopolymer blends. We consider an apparent interaction parameter 
$\chi_{a}$ that is defined by applying the RPA to the small $k$ limit 
of $S(k)$.  The predicted deviation of $\chi_{a}$ from its long chain limit is 
proportional to $N^{-1/2}$, where $N$ is chain length. This deviation is 
positive (i.e., destabilizing) for weakly non-ideal mixtures, with 
$\chi_{a} N \alt 1$, but negative (stabilizing) near the critical point. 
The positive correction to $\chi_{a}$ for low values of $\chi_{a} N$ is a
result of the fact that monomers in mixtures of shorter chains are slightly
less strongly shielded from intermolecular contacts.  The depression in 
$\chi_{a}$ near the critical point is a result of long-wavelength 
composition fluctuations. The one-loop theory predicts a shift in the 
critical temperature of ${\cal O}(N^{-1/2})$, which is much greater than 
the predicted ${\cal O}(N^{-1})$ width of the Ginzburg region. Chain 
dimensions deviate slightly from those of a random walk even in a 
one-component melt, and contract slightly with increasing $\chi_{e}$. 
Predictions for $S(k)$ and single-chain properties are compared to 
published lattice Monte Carlo simulations. 
\end{abstract}

\maketitle

\section{Introduction}
\label{sec:Intro}

Equilibrium properties of polymer mixtures that involve correlations
over distances greater than a few Angstroms are rather well described 
by a family of closely related self-consistent field theories.  Among 
these are Flory-Huggins (FH) theory for the free energy of a homogeneous 
mixture and the random phase approximation (RPA) theory of composition 
fluctuations. Both of these are special cases of a more general 
self-consistent field theory (SCFT) of inhomogenous polymer liquids.

Theoretical treatments of polymer mixtures that go beyond SCFT may be 
roughly classified as either microscopic liquid state theories 
or coarse-grained theories of correlation effects.
Liquid state theories such as the polymer reference interaction site 
model (PRISM) 
\cite{Schweizer1994}
and 
Born-Green-Yvon theories 
\cite{Lipson_98}
are designed to treat models with strong repulsive interactions, and 
attempt to predict all details of liquid structure from first principles. 
Coarse-grained theories are based on more idealized models of chain 
conformations and interactions, e.g., as continuous Gaussian chains 
with point-like interactions, similar to those used in coarse-grained
SCFT.

Over the past 20 years, a 
series of coarse-grained theories of correlation effects in 
polymer mixtures have made use of a functional integral 
representation of the partition function that was originally 
introduced by Edwards.
\cite{Cruz_Edwards_88,Bates_Fredrickson_94,Fredrickson_Liu_94,
Holyst_Vilgis_93,Holyst_Vilgis_94,Wang_95,Wang_02} 
All of these "field theoretic" studies of mixtures have thus 
far used an approximation in which the distribution of field 
fluctuations is approximated by a Gaussian functional.  This 
approximation is referred to in the jargon of field theory 
(which we have adopted) as a ``one-loop" approximation. An
alternative way of deriving this theory, which we find more
informative, is discussed in Sec. (\ref{sec:ThermoInt}) of 
this paper.

\subsection{Coarse-Graining and Renormalization}
The reason for studying correlations in coarse-grained
models is the hope that, by doing so, one might describe
phenomena that are not captured by SCFT, but that are, in 
some sense, independent of short-wavelength details. The 
first several studies of coarse-grained models of blends
\cite{Cruz_Edwards_88,Bates_Fredrickson_94,Fredrickson_Liu_94,
Holyst_Vilgis_93,Holyst_Vilgis_94,Wang_95} were, however, 
plagued by a sensitivity of all predictions to the treatment 
of very short wavelength correlations, at length scales which
the coarse-grained Hamiltonian was never intended to describe. 
The way around this problem, which was first applied in this 
context by Wang \cite{Wang_02}, is a renormalization procedure 
in which the sensitivity to monomer scale structure is 
absorbed into the values of a few phenomenological parameters. 
Wang \cite{Wang_02} considered the one-loop prediction for the 
free energy of a homogeneous polymer blend, and showed that the 
sensitivity of this quantity to local fluid structure could be 
absorbed into an appropriate renormalization of the Flory-Huggins 
interaction parameter.  More recently, we \cite{Grzywacz_Morse} 
showed how a generalization of this procedure can be used to 
obtain unambiguous results for one-loop corrections to both 
collective and single-chain correlation functions at nonzero 
wavenumbers, using the same formalism for polymer blends and 
diblock copolymer melts. 

Development of this renormalization procedure required two 
insights that were absent from the interpretation of earlier 
studies of the one-loop theory:

The first was that the SCF approximation should be identified
with the infinite chain limit of the one-loop theory, rather
than a simple random-mixing approximation.  Prior to Wang's 
work, the SCF approximation had often been conceptually 
identified with a simple mean-field approximation that is 
obtained from a saddle-point approximation for the underlying
functional integral. This mean-field theory ignores all 
correlations except those arising from intramolecular 
connectivity, which is never a quantitatively useful approximation. 
Defining SCF parameters by a process of extrapolation to the 
infinite chain limit instead takes advantage of the fact that
local correlations in liquids of long polymers depend only 
weakly on chain length, even in strongly correlated liquids.
The difference between the free energy of a system of finite 
chains and that of a hypothetical reference system of infinite 
chains is largely the result of slight differences in 
long-wavelength correlations. These differences may be adequately 
described by a coarse-grained model. By treating the interaction 
free energy of an infinite-chain reference system as an input to 
the theory (which must be inferred from experimental or simulation 
data), one can obtain a renormalized theory that makes unambiguous 
predictions of corrections to SCF phenomenology.

A second necessary insight was that the one-loop prediction for 
the structure factor $S(k)$ at nonzero wavenumber $k$ is affected 
by a renormalization of statistical segment length, as well as 
of the interaction free energy. The value of the statistical 
segment length $b$ of a flexible polymer in a dense mixture is 
generally different than the ``bare" value required as an input 
to the theory. The one-loop predictions for the magnitude of 
this difference \cite{Wang_95}, are sensitive to short-wavelength
correlations. This renormalization of $b$ has no effect upon the 
predicted value of structure factor $S(k)$ at exactly $k=0$, and 
so did not need to be taken into account in Wang's interpretation
of one-loop predictions for $S(k=0)$. It becomes an unavoidable 
complication, however, in calculations of single-chain correlations, 
and in applications to diblock copolymer melts, for which the 
dominant fluctuations occur at nonzero $k$.  

In our previous paper \cite{Grzywacz_Morse}, we focused primarily 
on clarifying the physical reasoning underlying our renormalization 
procedure, and on demonstrating that it yields a physically consistent 
interpretation. In this paper, we present predictions for binary 
homopolymer blends.  Predictions for diblock copolymer melts will 
be presented elsewhere. 

\subsection{Related Prior Work}
The analysis given here has several particularly close antecedents.

Our analysis of the free-energy density of a homogeneous blend and
the long wavelength limit of $S(k=0)$, which is given in Secs. 
\ref{sec:Theory} and \ref{sec:Collective}, is closely related to 
that of Wang.\cite{Wang_02} The main differences arise from 
mathematical approximations that Wang introduced in order to 
obtain a completely analytic approximation for behavior near the 
spinodal.  We have avoided all approximations other than the 
one-loop approximation itself, at the cost of some additional 
numerical analysis. The additional rigor has allowed us to calculate 
some additional quantities (e.g., the absolute shift in the 
critical point) and to identify physical phenomena that are 
important far from the spinodal.

Our analysis of corrections to Gaussian chain statistics in blends,
given in Sec. \ref{sec:SingleChain}, is closely related to the recent 
theoretical analysis of chain statistics in one-component melts by 
Beckrich, Johner, Semenov, and coworkers \cite{Beckrich_07, Wittmer_07a}. 
Like us, but prior to our work, these authors introduced a renormalized 
statistical segment length, defined by extrapolation to $N = \infty$, 
as part of a systematic calculation of small corrections to random-walk 
statistics.  We initially failed to notice the connection to our own 
work, \cite{Grzywacz_Morse} or to cite their work. These authors 
presented a beautiful analysis of universal corrections to Gaussian 
statistics in an incompressible melt at intermediate wavelengths, 
less than the coil size but greater than a monomer size. They also 
obtained analytic results for one-loop corrections to the intramolecular 
correlation function at all wavenumbers in melts of equilibrium 
polymers, with an exponential distribution of chain lengths. By
introducing a numerical procedure for extracting UV convergent parts 
of divergent integrals, we are also able to numerically evaluate 
one-loop predictions for melts and blends of monodisperse homopolymers 
and (elsewhere) for block copolymer melts. Here, we consider 
statistics of monodisperse polymers in binary blends.

The most important antecedent of both our work and the recent work 
of the Strasbourg group is the much larger, and (generally) more 
sophisticated theoretical literature on correlations in semi-dilute 
polymer solutions.\cite{DesCloizeaux1990} At a technical level, the 
closest analog of our approach within this literature seems to 
be the analysis of mono-disperse semi-dilute solutions by Ohta 
and Nakanishi.\cite{Ohta_Nakanishi_83,Ohta_Nakanishi_85} 

\subsection{Outline}
The paper is organized as follows: Sec. II provides a brief review of the
renormalized one-loop theory for the free energy density, and of our notation.
Sec. III presents predictions for corrections to the RPA theory of 
composition fluctuations. Sec.  IV presents predictions for corrections to
random walk statistics in polymer blends. Sec. V presents comparisons of 
theoretical predictions to results of published lattice Monte Carlo simulations.  
Sec. VI discusses the physical relationship between the one-loop theory 
for the free energy and RPA theory of correlations.  Sec. VII clarifies 
the physical origin of some features of the one-loop theory by relating 
them to corresponding features of the underlying RPA theory.

\section{Theoretical Formulation}
\label{sec:Theory}
\subsection{Model and Notation}
\label{sub:Model}
We consider a coarse-grained model of a binary homopolymer blend. Let $v$
be the average volume per monomer, for either species.  Let the volume 
fraction, degree of polymerization, statistical segment length and packing 
length \cite{Fetters_94} for species $i$ be denoted by $\phi_i$, $N_i$, 
$b_i$, and $l_{i} \equiv v/b_{i}^{2}$, respectively, for species $i=1$ 
and 2.  

Let the potential energy be the sum of an intramolecular potential 
$U_\text{chain}$ and a monomer-monomer pair interaction 
\begin{equation}
   U_{\rm int}
   = \frac{1}{2} \int d\rv \int d\rv'\ U_{ij}(\rv-\rv')c_i(\rv)c_j(\rv')
   \quad. \label{Uintdef}
\end{equation}
Here, $c_i(\rv) = \sum_{ms}^i \delta(\rv - \Rv_{mi}(s))$ is the concentration 
of monomers of type $i$, in which $\Rv_{mi}(s)$ is the position of monomer $s$ 
on molecule $m$ of species $i$.  The strength of the pair interaction is
taken to be of the form
\begin{equation}
   U_{ij}(\rv-\rv') = v 
   \begin{pmatrix} 
   B & B+\chi_0 \\
   B+\chi_0 & B
   \end{pmatrix}
   \delta_{\Lambda}(\rv-\rv') 
   \quad. \label{Uijmatrix}
\end{equation}
Here, $\delta_\Lambda$ is a function 
$\delta_\Lambda \equiv F \left( \Lambda |\rv - \rv'| \right) \Lambda^{-3}$ 
with a range $\Lambda^{-1}$ and an integral $\int d\rv u_\Lambda = 1$, so 
that $\delta_{\Lambda}$ approaches a Dirac $\delta$-function as $\Lambda 
\rightarrow \infty$. Here, $B$ is a dimensionless compression energy and 
$\chi_{0}$ is a dimensionless ``bare" interaction parameter. 
The incompressible limit may obtained by considering the behavior when $B$ 
is very large.

Let 
\begin{equation}
   S_{ij}(\rv) = \langle \delta c_i(\rv) \delta c_j(0) \rangle
   \label{Sijdef}
\end{equation}
where $\delta c_{i}(\rv) = c_{i}(\rv) - \langle c_{i} \rangle$, and 
let $S_{ij}(\kv) = \int d\rv \; e^{i\kv\cdot\rv} S_{ij}(\rv)$ denote
the Fourier transform.  Let $\Omega_{i}(\kv)$ be the intramolecular 
contribution to $S_{ii}(\kv)$ from chains of type $i$.  A continuous 
random walk model yields
\begin{equation}
   \Ssh_{i}(\kv) = \phi_{i}N \Debye(k^{2}R_{g,i}^{2})/v
   \label{SshDebye}
\end{equation}
where $R^{2}_{g,i} = N_{i}b_{i}^{2}/6$ is the radius of gyration 
for species $i$ and $\Debye(x) = 2(e^{-x}-1+x)/x^{2}$ is the Debye 
function.

Composition fluctuations in an effectively incompressible blend may 
be described by a scalar correlation function 
\begin{equation}
   S(k) = S_{11}(k) = S_{22}(k) = - S_{12}(k)
   \quad. \label{SscalarDef}
\end{equation}
Here, we follow Schweizer and Curro and our previous work, by 
expressing this scalar correlation function in terms of what we
(and Wang \cite{Wang_02}) refer to as an ``apparent" interaction 
parameter $\chi_{a}(k)$ , by defining
\begin{equation}
  S^{-1}(k) = \Ssh^{-1}_{1}(k) + \Ssh^{-1}_{2}(k) - 2v\chi_{a}(k)
  \quad. \label{chiaq_def}
\end{equation}
This is, of course, a generalization of the RPA expression for
$S(k)$, in which it is assumed that chains are random walks and 
that $\chi_{a}(k)$ is independent of $k$, $N$, and $\chi N$.

\subsection{One-loop Free Energy}
\label{sub:FreeEnergy}
The Helmholtz free energy $f$ per volume for this coarse-grained model 
may be expressed, in the incompressible limit, as a sum
\begin{equation}
   f = f_{\rm id} + e_{\rm mf} 
   + f_{\rm corr}
   \quad, \label{f_tot}
\end{equation}
where 
\begin{equation}
   f_{\rm id} \equiv \frac{\phi_1}{N_1v}\ln \phi_1 + \frac{\phi_2}{N_2v}\ln \phi_2 
   \quad. \label{FHideal}
\end{equation}
is the ideal Flory-Huggins entropy of mixing, and
\begin{eqnarray}
   e_{\rm mf} & \equiv & \frac{1}{2} \int d\rv \; U_{ij}(\rv)
   \langle c_{i} \rangle \langle c_j \rangle 
   \nonumber \\ 
   & = &
   v^{-1} \chi_0 \phi_{1}\phi_{2}
\end{eqnarray}
is a simple mean-field approximation for the interaction energy 
density.  The remainder, denoted $f_{\rm corr}$, is a correlation 
free energy. Here, and throughout the remainder of this paper
except Sec. V, where we compare to simulation results, we set
$kT=1$.

A one-loop approximation for $f_{\rm corr}$ has been obtained in 
previous work by considering a Gaussian approximation for the 
distribution of fluctuations of a complex chemical potential field 
within the context of the Edwards auxiliary field formalism. In 
Sec. \ref{sec:ThermoInt}, we discuss an alternative derivation in 
which this approximation is instead obtained by thermodynamic 
integration of the RPA correlation function. However it is obtained, 
the resulting approximation for $f_{\rm corr}$ is given, in the 
general case of a compressible liquid, by a Fourier integral
\begin{equation}
    f_{\rm corr} = 
    \frac{1}{2} \int_{\qv} 
    \ln \left [  {\rm det}| \Iv + \Sshb(\qv) \Uhb(\qv) | \right ]
     \label{dfOneLoopBare0}
\end{equation}
Here, the bold faced symbols $\Sshb(\qv)$ and $\Uhb(\kv)$ indicate 
$2\times 2$ matrices with elements $\delta_{ij}\Omega_{i}(\qv)$ and 
$U_{ij}(\qv)$, respectively, $\Iv$ denotes the identity, and 
${\rm det}| \cdots |$ is a determinant. Here and hereafter, we use 
the shorthand $\int_{\qv} \equiv \int d^{3}q/(2\pi)^{3}$ for Fourier 
integrals. 

In the nearly-incompressible limit of interest here, 
Eq. \ref{dfOneLoopBare0} for $f_{\rm corr}$ may be approximately 
reduced to
\begin{eqnarray}
  f_{\rm corr}
  = \frac{1}{2} \int_{\qv} \ln \left[ \left(\Omega_1 
  + \Omega_2 - 2\chi_0 v \Omega_1 \Omega_2\right)v \right].
  \label{dfOneLoopBare}
\end{eqnarray}
To obtain the integrand of Eq. (\ref{dfOneLoopBare}),
in a model with a pair interaction of the form given in Eq. 
(\ref{Uijmatrix}), we have both taken the limit $B \gg 1$
in Eq. (\ref{dfOneLoopBare0}), and also replaced the Fourier 
transform of $\delta_{\Lambda}$ by unity.  This yields an 
expression for the integrand that is valid for $|\qv| \ll 
\Lambda$. 
In what follows, we regularize the one-loop approximation for 
$f_{\rm corr}$ by treating $\Lambda$ as a cutoff wavenumber, 
and simply restricting the integral in Eq. (\ref{dfOneLoopBare}) 
to $|\qv| < \Lambda$.  This is equivalent to the use of a pair 
potential in which the Fourier transform of 
$\delta_{\Lambda}(\rv-\rv')$ is taken to be unity for 
$\qv < \Lambda$, and zero for $\qv > \Lambda$. The purpose of 
the renormalization procedure discussed below is to extract 
predictions that are independent of such details of the treatment
of short-wavelength correlations.

\subsection{Divergence and Renormalization}
\label{sub:Renormalization}
It has long been known that, in a model of continuous Gaussian 
chains, the idealized limit $\Lambda \rightarrow \infty$ of 
point-like interactions yields a one-loop expression for 
$f_{\rm corr}$ that diverges as the cube of $\Lambda$.
\cite{Cruz_Edwards_88, Bates_Fredrickson_94, Fredrickson_Liu_94, Wang_02}, 
In previous work, both Wang \cite{Wang_02} and we \cite{Grzywacz_Morse} 
have analyzed the structure of the high-$q$ divergence of this free 
energy density, and its physical significance.
We found that Eq. (\ref{dfOneLoopBare}) for $f_{\rm corr}$ can be
expressed as a sum
\begin{equation}
   f_{\rm corr} = f^{\Lambda}_{\rm corr} + f^{*}
   \label{f_Lambda_star}
\end{equation}
of an ``ultra-violet" (UV) divergent contribution 
$f^{\Lambda}_{\rm corr}$, which increases with increasing $\Lambda$, 
and a contribution $f^{*}$ that is independent of $\Lambda$.  The UV 
divergent contribution $f_{\rm corr}^{\Lambda}$ may be expressed as
a sum 
\cite{Grzywacz_Morse} 
\begin{equation}
   f^\Lambda_{\rm corr} = f^\Lambda_{\rm bulk} + f^{\Lambda}_{\rm end}
\end{equation}
where
\begin{eqnarray}
   f^{\Lambda}_{\rm bulk}
   & = & 
   \frac{1}{12\pi^2}
   \left[ \ln\left( \frac{12\bar{l}}{v\Lambda^2} \right) + \frac{2}{3} \right] \Lambda^3
   -\frac{6\chi_0} {\pi^2v} \frac{l_1 l_2}{\bar{l}} \phi_1\phi_2\Lambda  
   \nonumber\\
   f^{\Lambda}_{\rm end} & = &
   -\frac{3}{2\pi^2v\bar{l}}
   \left( \frac{\phi_1 l_1^2}{N_1} + \frac{\phi_2 l_2^2}{N_2} \right) \Lambda .
   \label{GaussianUV}
\end{eqnarray}
and where $\bar l \equiv \phi_1 l_1 + \phi_2 l_2$. The contribution 
$f^\Lambda_{\rm end}$, which is proportional to $1/N$, was identified 
\cite{Grzywacz_Morse} as an excess free energy for chain ends. The 
quantity $f^{\Lambda}_{\rm bulk}$, which is independent of $N$, is 
instead a correction to the ``bulk" interaction free energy density 
of a hypothetical reference system of infinitely long chains. 

To remove the explicit cutoff dependence, we re-interpret the one-loop 
free energy (excluding end effects) as a sum
\begin{equation}
   f = f_{\rm id} + f_{\rm int} + f_{\rm end}^{\Lambda} + f^{*}
   \quad. \label{f_id_int_star}
\end{equation}
in which
\begin{equation}
    f_{\rm int} \equiv 
    e_{\rm mf}
    + f^{\Lambda}_{\rm bulk}
    \quad.  \label{f_int_def}
\end{equation}
is identified as the interaction free energy of the relevant form
of SCFT. The contribution $f^{\Lambda}_{\rm end}$ is a non-universal
chain end contribution that could (and generally should) be added to 
SCFT, to reflect the fact that the fluid structure is perturbed in 
the immediate vicinity of any chain end.  The remaining 
cutoff-independent contribution $f^{*}$, is identified as a 
universal correction to the SCF phenomenology. 

Eq. (\ref{f_int_def}) for $f_{\rm int}$ simplifies considerably 
in the case of structurally symmetric polymers, with $l_{1}=l_{2} = l$. 
In this case $f_{\rm int}$ is given, to within a composition-independent 
constant, by a function of the form
\begin{equation}
   f_{\rm int} = v^{-1} \chi_e(\Lambda) \phi_1\phi_2 
   \quad.  \label{f_int_sym}
\end{equation}
where 
\begin{equation}
   \chi_e(\Lambda) = \left [ 1 - \frac{6}{\pi^{2}} l \Lambda \right ] \chi_{0}
   \quad.  \label{chie_sym}
\end{equation}
is an effective Flory-Huggins interaction parameter. This UV 
divergent expression for $\chi_e$ was first obtained by Olvera de la 
Cruz {\it et al.} \cite{Cruz_Edwards_88}.  The physical origin of the 
UV divergence of this quantity is discussed in Sec. \ref{sec:CorrHole}.

We focus in the remainder of this paper on the UV convergent correction 
$f^{*}$. It is convenient to introduce a notation for $f^{*}$ and 
related quantities, in which we use $\int_{\qv}^{*}$ to denote a 
renormalized Fourier integral. The renormalized Fourier integral of 
an integrand $A(\qv)$ is defined by a difference
\begin{equation}
   \int_{\qv}^{*} A(\qv) \equiv \int_{\qv} A(\qv) 
                              - \int_{\qv}^{\Lambda} A(\qv)
   \label{int_star_def}
\end{equation}
between the unrenormalized integral $\int_{\qv} A(\qv)$ and the UV 
divergent part of the integral, denoted by $\int_{\qv}^{\Lambda}A(\qv)$.
In a theory of continuous chains, the ``UV divergent" part of Fourier 
integral is obtained by constructing a high-$q$ asymptotic expansion 
of the integrand, in decreasing powers of $qR$, and retaining only 
those terms in the expansion that lead to a UV divergent integral. 
\cite{Grzywacz_Morse}

With this notation, we may write
\begin{eqnarray}
    f^{*} = \frac{1}{2} \int_{\qv}^{*} 
    \ln \left[ \left(\Omega_1 + \Omega_2 
    - 2 v \chi \Omega_1 \Omega_2\right)v \right].
    \label{dfOneLoop}
\end{eqnarray}
The $N$-dependence of $f^{*}$ may be isolated by non-dimensionalizing
the renormalized integral in terms of a reference length $R$, which we 
take to be the root-mean-squared end-to-end $R = R_{g1}$ of species 1.  
It may be shown that all of the quantities in the integrand can then 
be expressed as functions of $qR$ and of a set of dimensionless parameters 
$\phi_{1}$, $\chi N$, $N_{2}/N_{1}$, $R_{2}/R_{1}$.  By this method, 
we find that $f^{*}$ may be expressed as a function of the form
\begin{equation}
   f^{*} = \frac{1}{R^{3}}
  \hat{f}^{*}(\chi N, \phi_{1}, N_{2}/N_{1}, R_{2}/R_{1})
  \label{df_nondim}
\end{equation}
in which $\hat{f}^{*}$ is a dimensionless function given by the value 
of the non-dimensionalized renormalized integral. The required set 
of dimensionless parameters is the same as that required by the 
standard non-dimensionalized SCFT for an incompressible blend.
Note that, in Eqs.  (\ref{dfOneLoop}) and (\ref{df_nondim}), we have 
written $f_{\rm corr}$ as a product of a parameter $\chi$, but have
not specified whether by $\chi$ we mean $\chi_{0}$, $\chi_{e}$, or 
(perhaps) the low-$q$ limit of $\chi_{a}$. In fact, different choices
lead to different variants of the theory, as discussed in more detail
below.

Results for $f^{*}$ and related quantities that are reported here have 
all been obtained by numerically evaluating the relevant renormalized 
Fourier integrals. To do this, for each quantity of interest, we first 
numerically evaluate the unrenormalized integral with a large but finite 
cutoff $|\qv| < \Lambda$, and then subtract an analytic result for the
UV divergent part of the regularized integral.  The Fourier integrals 
required by the one-loop theory may all be expressed as either one 
dimensional integrals with respect to a wavenumber $q$ (for free energy
density and its derivatives) or two-dimensional integrals with respect 
to $q$ and a polar angle $\theta$  (for correlation functions at 
nonzero wavenumber).  Subtraction of the UV 
divergent part from an unrenormalized integral yields an estimate of 
the renormalized integral that depends only weakly on the numerical 
cutoff $\Lambda$.  Our final result for the value of a renormalized 
integral is obtained by repeating this procedure for several values of 
$\Lambda \gg R$, and numerically extrapolating to $\Lambda = \infty$. 

\subsection{Structure Function}\label{chi}
\label{sub:Collective}
The long-wavelength limit of $S(k)$ is related to the composition 
dependence of the free energy density $f$ by a statistical mechanical 
theorem:
\begin{equation}
   \lim_{k \rightarrow 0}S^{-1}(k) 
   = v^{2} \frac {\partial^2 f} {\partial \phi_1^2}
   \quad.
\end{equation}
The quantity $S^{-1}(0)$ is also related to the long-wavelength limit 
of $\chi_{a}(k)$ by
\begin{equation}
  \lim_{k \rightarrow 0} S^{-1}(k) =
  2v(\chi_{s} - \chi_{a}) \quad.
\end{equation}
Here, and hereafter, we use $\chi_{a}$ to denote the $\qv \rightarrow 0$
limit of the quantity $\chi_{a}(\kv)$ defined in Eq. (\ref{chiaq_def}). 
The quantity $\chi_{s}$ is the spinodal value of $\chi_{a}$, given by
\begin{equation}
  2\chi_{s} \equiv 
  v \frac {\partial^2 f_{\rm id}} {\partial \phi_1^2} =
  \frac{1}{N_{1}\phi_{1}} + \frac{1}{N_{2}\phi_{2}}
\end{equation}
Using decomposition (\ref{f_id_int_star}) for $f$, and hereafter 
neglecting the explicit end effects, the one-loop approximation 
for $\chi_{a}$ may be expressed as a sum
\begin{equation}
    \chi_a = \chi_e + \chi^{*}
    \quad,
\end{equation}
in which
\begin{equation}
   \chi_e \equiv
   -\frac{v}{2} \frac {\partial^2 f_{\rm int}} {\partial \phi_1^2}
   \label{chie_def}
\end{equation}
is the interaction parameter of phenomenological SCFT, and
\begin{equation}
    \chi^{*} = -\frac{v}{2} 
    \frac {\partial^2 f^{*}} {\partial \phi_1^2}
\end{equation}
is a correction to SCFT arising from long-wavelength fluctuations. 

An explicit Fourier integral expression for $\chi^{*}$ may be obtained 
by differentiating the integrand of Eq. (\ref{dfOneLoop}) for $f^{*}$ 
twice with respect to $\phi_{1}$. The resulting approximation for
$\chi^{*}$ may be expressed in terms of either Edwards' screened 
interaction \cite{Grzywacz_Morse} or in terms of the correlation 
function
\begin{equation}
   S^{-1}(k;\chi) = \Omega_{1}^{-1}(k) + \Omega_{2}^{-1}(k) - 2v\chi
   \quad.
\end{equation}
When expressed in terms of $S(k)$, it is found that
\begin{eqnarray}
   \chi^{*} = 
   \frac{v}{4} \int_{\qv}^{*}
   \left [  \mu^{2}(q) S^{2}(q;\chi) - \lambda(q)S(q;\chi) + \eta \right ]
   \label{chia_correction}
\end{eqnarray}
where
\begin{eqnarray}
   \mu(q) &=&  
   \frac{1}{\phi_{1}\Omega_{1}(q)} 
   - \frac{1}{\phi_{2}\Omega_{2}(q)}
   \nonumber\\
   \lambda(q) &=& -2 \left( 
           \frac{1}{\phi_{1}^{2}\Omega_{1}(q)} + 
           \frac{1}{\phi_{2}^{2}\Omega_{2}(q)} \right ) 
\nonumber \\
\eta &=&  1/\phi_1^2 + 1/\phi_2^2.
\end{eqnarray}
The quantities $\mu(0)$ and $\lambda(0)$ are proportional to 
the third and fourth derivatives of $f_{\rm id}(\phi_{1})$ 
with respect to $\phi_{1}$, respectively.  Since $\mu(q)$, 
$\lambda(q)$ and $\eta$ are independent of $\chi$, the 
proximity to the spinodal in this expression is controlled 
by $S(q)$ alone. 

By non-dimensionalizing the integral in Eq. (\ref{chia_correction}),
we may show that it yields a correction to $N \chi_{a}$ of the  
form
\begin{equation}
   N\chi^{*} = \frac{1}{\bar{N}^{1/2}}
   \hat{\chi}^{*}(\chi N, \phi_{1}, N_{2}/N_{1}, R_{g2}/R_{g1})
   \quad, \label{dchi_nondim}
\end{equation}
in which $\hat{\chi}^{*}$ is a non-dimensional function that is 
defined by the non-dimensionalized integral. Here,
\begin{equation}
   \bar{N} \equiv Nb^{6}/v^{2}
\end{equation}
is an invariant degree of polymerization, in which (by convention) 
$N=N_1$ and $b=b_1$. 

In Eqs. (\ref{dfOneLoop}), (\ref{df_nondim}), and (\ref{dchi_nondim}),
we have intentionally expressed $f^{*}$ and $\chi^{*}$ as functions of
an interaction parameter $\chi$, without specifying whether this input
parameter should be taken to be the bare parameter $\chi_{0}$ or some 
type of a renormalized value.  In fact, different variants of the 
renormalized one-loop theory can be obtained by different choices for 
the interaction parameter used within these Fourier integrals.
If we define $f^{*}$ by simply subtracting the UV divergent part of 
the one-loop expression for $f_{\rm corr}$, we obtain a theory that 
is no longer UV divergent, but that retains an explicit dependence 
upon the unmeasurable ``bare" parameter $\chi_{0}$. Following the
logic used to construct renormalized expansions of field theories, 
however, we may also replace $\chi_{0}$ by a renormalized parameter 
within each integrand to obtain a theory from which all reference 
to the bare parameter have been removed. This substitution may be 
formally justified, within the context of a one-loop approximation, 
by observing that the difference between approximations obtained by 
using a bare or renormalized parameter in the one-loop theory differ 
only at second order in a systematic loop expansion. 

In the remainder of this paper, we follow Wang \cite{Wang_02}
by considering a self-consistent one-loop approximation. In this
approximation, also known as a ``Hartree" approximation, $\chi$
is replaced within the Fourier integral expression for $\chi^{*}$ 
by $\chi_{a} \equiv \chi_{a}(\qv=0)$.  In the resulting theory,
$\chi_{a}$ is given by a function of the form
\begin{equation}
    \chi_{a} = \chi_{e} + \chi^{*}(\phi_1, \chi_{a}) 
    \quad. \label{chia_of_chia}
\end{equation}
Here, $\chi_{e}$ is related to the SCF (i.e., $N \rightarrow \infty$)
interaction free energy $f_{\rm int}(\phi)$ by Eq. (\ref{chie_def}).
For purposes of numerical evaluation, it is convenient to rewrite 
Eq.  (\ref{chia_of_chia}) as an explicit expression 
$\chi_{e} = \chi_{a} - \chi^{*}(\phi_1, \chi_{a})$
for $\chi_{e}$ as an explicit function of $\chi_{a}$. 

\section{Collective Fluctuations} 
\label{sec:Collective}

\subsection{Symmetric Blends, Critical Composition}
\label{subsec:Critical}

\begin{figure}[htb]\center
\includegraphics[width=0.40\textwidth,height=!]{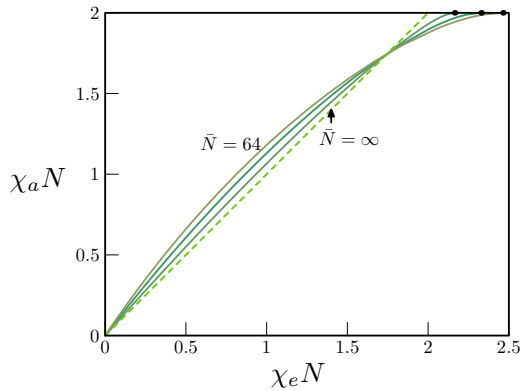}
\caption[$\chi_a(0)N$ versus $\chi_eN$]
{Self-consistently calculated $\chi_a N$ versus $\chi_e N$ for 
symmetric binary blends with $\phi_{1}=1/2$ for $\bar N$ = 64, 
128, 512 respectively (solid lines).  The dotted line shows
where $N\chi_{a}=N\chi_{e}$, which corresponds to the limit 
$\bar{N} \rightarrow \infty$. Solid dots along the line 
$\chi_{a}N=2$ indicate critical points for finite chains.
} 
\label{chiae_k0}
\end{figure}

Figs. \ref{chiae_k0} displays predictions of the renormalized 
Hartree theory for $\chi_{a}$ in blends of structurally symmetric 
polymers ($b=b_{1}=b_{2}$ and $N=N_{1}=N_{2}$) at their critical 
composition ($\phi_{1}=1/2$). It shows predictions 
for $\chi_a N$ as a function of $\chi_e N$ blends with three 
different chain lengths, corresponding to $\bar{N}$ = 64, 128, 
and 512. The difference between $\chi_{a}N$ from $\chi_{e}N$ is 
largest for the shortest chains, and vanishes in the limit 
$N \rightarrow \infty$.  The SCF predicts a critical point for 
a symmetric $\chi_{e}N=2$. The actual critical point for such 
a blend is reached when $\chi_{a}N = 2$. 

\begin{figure}[htb]\center
\includegraphics[width=0.40\textwidth,height=!]{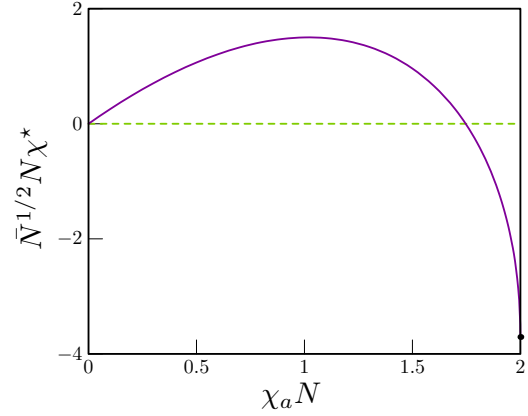}
\caption[$\sqrt{\bar N} N \chi_{*} $ versus $\chi_e N$]
{The deviation $\bar{N}^{1/2}N (\chi_{a} - \chi_{e})$ as a
function of $\chi{a} N$ for symmetric binary blends at the 
critical composition, $\phi_{1}=1/2$.}
\label{dchiN}
\end{figure}

Fig. \ref{dchiN} shows the underlying dimensionless function 
$\hat{\chi}^{*} \equiv (\chi_{a}N - \chi_{e}N) \sqrt{\bar{N}}$
as a function of $\chi_{a}N$. Note that both $\chi_{a}$ and 
$\chi^{*}$ vanish in the limit $\chi_{e}=0$ for any structurally 
symmetric mixture, with $b_{1}=b_{2}$ and $N_{1}=N_{2}$, because 
this is an exact result for such a ideal mixture, in which the
two species are physically indistinguishable. 

Notably, the dependence of $\chi^{*}$ upon $\chi_{a} N$ is
not monotonic: For small values of $\chi_{a} N$, the deviation 
$\chi^{*}=\chi_{a}-\chi_{e}$ is positive and increases with 
increasing $\chi_{a}$.  Closer to the critical point, $\chi^{*}$ 
begins to decrease with increasing $\chi_{a}N$, changes sign 
at $\chi_{a}N \simeq 1.7$, and is negative at the critical 
point. Note that all the curves in Fig. (\ref{chiae_k0}) 
intersect at $\chi_{a}N = \chi_{e}N \simeq 1.7$, where $\chi^{*}=0$.  
For any structurally symmetric blend, with $b_{1}=b_{2}$ and 
$N_{1}=N_{2}$, we find that the initial slope is 
\begin{eqnarray}
   \lim_{\chi_a N \rightarrow 0} 
   \frac{\partial ( \chi^{*} N) }{\partial ( \chi_{a} N)} = 
   \frac{(6/\pi)^{3/2}}{\bar{N}^{1/2}} , 
   \label{corrl_hole}
\end{eqnarray}
for any composition.  

At the critical point of such a blend ($\phi_{1}=1/2$ and 
$\chi_{a}N = 2$), the one-loop prediction for $\chi^{*}N$ 
approaches a negative limiting value
\begin{equation}
   \lim_{ \chi_{a} N \rightarrow 2} \chi^{*} N
   = - 3.7\bar{N}^{-1/2} \quad,
\end{equation}
with a slope that diverges at the critical point. In the 
Hartree approximation, this yields a predicted critical
value of
\begin{equation}
   (\chi_e N)_c = 2 + 3.7\bar{N}^{-1/2}. 
   \label{critical_shift}
\end{equation}
If $\chi_{e}$ is a decreasing function of temperature, this
corresponds to a depression in the critical temperature 
$T_{c}$ by an amount $\delta T_{c} \propto N^{-1/2}$ 

The predicted decrease in $\chi^{*}$ near the critical point is 
not surprising: Near the critical point, long wavelength 
correlations decrease the level of local mixing between $A$ 
and $B$ monomers, thereby decreasing the apparent interaction 
parameter $\chi_{a}$ and stabilizing the homogeneous phase. 

The origin of the positive values obtained for $\chi^{*}$ at 
small values of $\chi_{a}$ is less obvious. In Sec. 
\ref{sec:CorrHole} of this paper we show that this prediction
is the result of a subtle dependence of local correlations in 
a polymer liquid upon overall chain length, which is discussed
in much greater detail in an accompanying paper. In a dense
liquid of long polymers, the immediate vicinity of each monomer 
is crowded with monomers that belong to the same chain. 
This intramolecular ``self-concentration" reduces the space 
available for monomers from other chains, thus creating a
so-called correlation hole in the intermolecular correlation 
function. In a mixture, the correlation hole reduces the 
magnitude of the potential energy of interaction between 
monomers of types $1$ and $2$. We show in the accompanying
paper that the depth of the correlation hole, decreases 
slightly with decreasing chain length. For from the spinodal,
this simple packing effect causes $\chi_{a}$ to increase 
with decreasing chain length, reflecting the fact that 
shorter chains are less strongly shielded from intermoecular
contacts. Because we define $\chi_{e}$ to be the limiting 
value of $\chi_{a}$ in the limit of infinitely long chains, 
this effect tends makes a positive contribution to the 
difference $\chi_{a}(N) - \chi_{e}$ for any finite $N$. 
In the accompanying paper, we analyze this effect without
reference to the one-loop theory, and derive the prefactor
in Eq.\ref{corrl_hole} by very different reasoning than 
that used here.

A shift in the critical temperature by an amount $\delta T_{c} 
\propto \bar{N}^{-1/2}$ was predicted previously by Holyst 
and Vilgis \cite{Holyst_Vilgis_93}. The one-loop theory of 
Holyst and Vilgis did not use a systematic renormalization
procedure, but instead introduced an {\it ad hoc} cutoff 
$\Lambda \sim 1/R$ to avoid the UV divergence. This approach 
captured the correct scaling, but does not allow a meaningful
calculation of numerical values for one-loop corrections, 
because it yields corrections whose values are sensitive 
to the exact numerical value chosen for an arbitrary cutoff 
length. 

\subsection{Vicinity of the Spinodal}
\label{sub:Spinodal}
We now discuss the behavior of the Hartree theory near the spinodal.
This part of our analysis is similar to that given by Wang
\cite{Wang_02}, who focused on behavior near the spinodal, but who 
made several approximations that we avoid. 

A rigorous asymptotic expansion of the behavior of $\chi_{a}$ near 
the spinodal may be obtained by expanding $S(q;\chi_{a})$, $\mu(q)$ 
and $\lambda(q)$ in Eq. (\ref{chia_correction}) around their values 
at $q=0$ and $\chi_{a} = \chi_{s}$. Let
\begin{eqnarray}
    r    & =  & 2(\chi_{s} - \chi_{a})
    \nonumber \\
    \tau & =  & 2(\chi_{s} - \chi_{e}).
\end{eqnarray}
The required asymptotic expansion of $S(q)$ is 
\begin{equation}
  S^{-1}(q) \simeq r + q^{2}\xi_{0}^{2}
  \quad. \label{S_expand}
\end{equation}
where 
\begin{equation}
 \xi_{0}^{2} \equiv v \bar{l}/(18\phi_1\phi_2 l_1 l_2)
 \quad.
\end{equation}
By also expanding $\mu(q)$ and $\lambda(q)$ around $q=0$, we obtain
an asymptotic expansion of the form
\begin{equation}
    r = \tau + \frac{v}{\xi_{0}^{3}N^{3/2}}
    \left [ \frac{A}{\sqrt{N r}} + B + C \sqrt{N r} + \cdots \right ]
\end{equation}
where $A$, $B$, and $C$ are dimensionless coefficients that are
independent of $r$, but that that depend upon $\phi_{1}$, $N_{2}/N_{1}$, 
and $R_{2}/R_{1}$. Here, $N$ is an arbitrary choice of a reference 
degree of polymerization, which could be taken to be $N_{1}$ or 
$N_{2}$ or an appropriate average of the two. 

The coefficients $A$ and $C$ in this expansion can be calculated
analytically, by considering the singular behavor of the integral
near $q=0$ as $r \rightarrow 0$. The coefficient $A$ is given
\begin{eqnarray}
    A(\phi_{1}) & = & 
    \frac{-N^{2}}{16\pi v^{2}} \mu^{2}(0)
    \nonumber \\
    & = & -\frac{N^{2}}{32\pi } \left ( 
    \frac{1}{\phi_{1}^{2}N_{1}} - \frac{1}{\phi_{2}^{2}N_{2}} 
    \right)^{2} \quad.
\end{eqnarray}
Here, $\mu(q=0)$ is the third derivative with respect to $\phi_{1}$ 
of the ideal mixture free energy per monomer $vf_{\rm id}(\phi_1)$. 
This quantity vanishes at the Flory-Huggins critical composition, 
at which $\phi_{1}^{2}N_{1} = \phi_{2}^{2}N_{2}$, and so (as noted
by Wang \cite{Wang_02}) $A(\phi_{1})$ also vanishes at the critical 
composition $\phi_{1c}$. At the critical composition, we find 
\begin{equation}
    C(\phi_{1c}) = \frac{-1}{4\pi \phi_{1c}^{2}\phi_{2c}^{2} }
    \frac{N}{ \sqrt{N_{1}N_{2}} }
    \quad.
\end{equation}
These values for $A$ and $C$ agree with corresponding results of
Wang\cite{Wang_02}. 

In the Hartree approximation, the value $\tau_{c}$ of $\tau$ 
(the reduced ``temperature") at the critical point is determined 
by the value of the constant $B$ at the critical composition: 
The critical point occurs when $r=0$ at $\phi_{1}=\phi_{1c}$,
or when
\begin{equation}
   \tau_{c} = - v B(\phi_{1c}) \xi_{0}^{-3}N^{-3/2}
\end{equation}
If $B$ is nonzero, an expansion of this form thus yields a shift 
in the critical value of $\chi_{e}N$ of order $\bar{N}^{-1/2}$.

The coefficient $B$ cannot be obtained from an asymptotic analysis of 
the low-$q$ behavior of the integral in the Eq.(\ref{chia_correction}).
To understand why, recall that the problem actually involves three 
length scales: the monomer size (or cutoff length $\Lambda^{-1}$), 
the coil size $R \propto \sqrt{N}$, and correlation length 
$\xi=\xi_{0}/\sqrt{r}$. Very near the spinodal, $\xi \gg R$. Our 
renormalization procedure removes all dependence on $\Lambda^{-1}$, 
but leaves a dependence on both $R$ and $\xi$. Away from the 
critical composition, the integral in Eq. (\ref{chia_correction}) 
develops an infra-red (IR) divergence of the form $A/\sqrt{r}$ as 
$r \rightarrow 0$.  This divergence is the result of fluctuations 
with wavenumbers $q \sim \xi^{-1}$, and so its prefactor can be 
obtained by considering the behavior of the integrand near $q=0$. 
At the critical composition, the coefficient $C$ can 
be isolated by considering the IR divergence of the derivative 
of Eq. (\ref{chia_correction}) 
$\chi^{*}$ with respect to $\chi_{a}$, which then diverges as 
$r^{-1/2}$. The actual value of $\chi^{*}$, however, convergences 
even at the critical point, and is dominated by fluctuations with 
$qR \sim 1$.  As a result, the one-loop prediction for the shift 
in the critical point can only be calculated by numerically 
evaluating the renormalized integral, as done here, without 
introducing approximations for $\Omega_{1}(q)$ and $\Omega_{2}(q)$ 
that are valid only for $qR \ll 1$.

At the critical composition, the asymptotic expansion of the 
Hartree theory may be expressed more compactly as a sum
\begin{equation}
   Nr = N\delta \tau  + \frac{Cv}{\xi_{0}^{3}N^{1/2}}\sqrt{Nr}
\end{equation}
where $\delta \tau \equiv \tau - \tau_{c}$, with $C < 0$. 
The theory exhibits strongly non-classical critical behavior 
for values of $\delta \tau$ less than a crossover value 
$\delta \tau^{*}$.  Approximating $\delta\tau^{*}$ by the 
value at which the $\sqrt{r}$ fluctuation correction is 
equal to $\delta\tau$ yields a reduced crossover temperature 
\begin{equation}
   N \delta \tau^{*} \sim \bar{N}^{-1}
   \quad,
\end{equation}
as first noted by de Gennes. \cite{deGennes_77}
As noted by Holyst and Vilgis \cite{Holyst_Vilgis_93}, 
the predicted ${\cal O}(1/N)$ with of the critical region 
is thus much less than the predicted ${\cal O}(N^{-1/2})$ 
magnitude of the shift of the critical temperature from 
its SCF value. 

Very close to the critical point, $\tau \ll \tau^{*}$, the 
Hartree theory yields $r \propto \tau^{2}$, or $S(q=0) \propto 
\tau^{-2}$. The predicted critical exponent of $\gamma = 2$ 
is much larger than the known Ising critical exponent of 
$\gamma \simeq 1.26$.  This is a well known defect of the 
Hartree theory.

\subsection{Off-Critical Blends}

\begin{figure}[htb]\center
\includegraphics[width=0.40\textwidth,height=!]{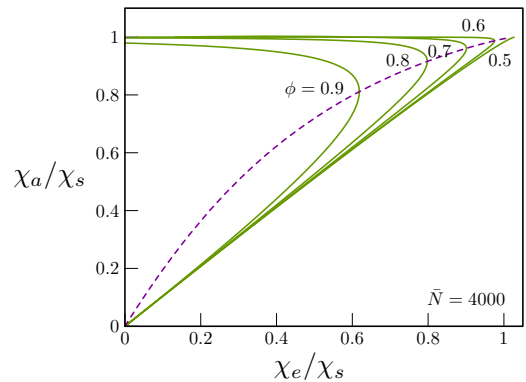}
\caption[Self-consistently computed $\chi_a$ versus $\chi_e$ 
for various compositions.]
{Self-consistently computed $\chi_a$ versus $\chi_e$ for 
off-critical blends with $b_1=b_2$ and $N_1=N_2$ and 
$\bar{N}=4000$ at various values of $\phi=\phi_{1}$. Both
axes are rescaled by the spinodal value $\chi_{s}$. Note
that the spinodal line $\chi_{a}/\chi_{s}=1$ is actually 
reached only at the critical composition of $\phi_{1}=1/2$. 
The computed locus of turning points is depicted with dashed 
line.
}
\label{offCritical}
\end{figure}

At any composition except the critical composition, the Hartree theory
considered here does not have a spinodal: At off-critical compositions, because
$\chi^{*}$ diverges like $r^{-1/2}$ with decreasing $r$, there is no value of
$\tau$ for which $r=0$. As noted by Wang \cite{Wang_02}, in the Hartree
approximation, this causes $\chi_{e}$ to become a non-monotonic function of
$\chi_{a}$ or (equivalently) $\chi_{a}$ to become a multi-valued function
$\chi_{e}$.  An example of this behavior is shown in Fig. \ref{offCritical},
which shows predictions of renormalized Hartree theory for $\chi_a$ vs.
$\chi_e$ in a series of off-critical blends of various compositions with $b_1 =
b_2$, $N_1 = N_2$, and $\bar{N} = 4000$.  The turning point at which $\chi_{e}$
reaches its maximum value, which is also where  $S(q=0)$ is maximum, is what
Wang referred to as a ``pseudo-spinodal".  We do not attach any physical
significance to this turning point, except as a particularly obvious failure of
the Hartree approximation.

i

\section{Single Chain Properties}
\label{sec:SingleChain}
The one-loop theory \cite{Morse_06, Grzywacz_Morse} also predicts 
slight deviations from random-walk statistics for individual chains. 
The one-loop approximation for the intramolecular correlations function 
is a form of renormalized perturbation theory in which each pair of 
monomers on a chain interact via Edwards' screened interaction, 
and in which the effect of this interaction is taken into account 
to first order in powers of the strength of the screened 
interaction\cite{Barrat_Fredrickson_91}.

We consider the single-chain correlation function $\omega_{i}(\kv)$. 
This is given for a Gaussian homopolymer by 
\begin{equation}
   \omega_{i}(\kv) = N^{2}g(k^{2}R_{g,i}^{2}) \quad,
\end{equation}
where $R_{g,i}=\sqrt{N_{i}b^{2}_{i}/6}$ is the radius of gyration
of species $i$ and $g(x) = 2(e^{-x}-1+x)/x^{2}$ is the Debye function.
The one-loop correction to $\omega_{i}(\kv)$ in a homopolymer blend 
is given by \cite{Morse_06, Grzywacz_Morse}
\begin{eqnarray}
   \delta\omega_i(\kv) = - \frac{1}{2}
   \int d\qv\ \psi_i^{(4)} (\kv, -\kv, \qv, -\qv) G_{ii}(\qv)
   \quad. \label{intra_corrl}
\end{eqnarray}
Here, $G_{ii}(\qv)$ is the screened interaction between any two $i$ 
monomers, which is given in an incompressible blend at wavelengths 
$q \ll \Lambda$ by
\begin{equation}
  G_{ii} = \frac{ 1 - 2 v \chi_e \Omega_1  \Omega_2  \Omega_i^{-1} }
                 { \Omega_1 + \Omega_2 - 2v \chi\Omega_1 \Omega_2 }
  \quad. \label{screened_interaction}
\end{equation}
The function $\psi_{i}^{(4)}=\psi_{i}^{(4)}(\kv,-\kv,\qv,-\qv)$ is 
given by
\begin{eqnarray*}
   \psi_i^{(4)}
   &=& \omega^{(4)}_i(\kv,-\kv,\qv,-\qv) -
   \omega_i(\kv) \omega_i(\qv), \\
\end{eqnarray*}
where $\omega^{(4)}_{i}$ is the 4-point intramolecular correlation 
function for an ideal Gaussian chain\cite{Grzywacz_Morse}. 
It is straightforward to verify that $\delta\omega_{i}(\kv)$ 
vanishes in the limit $\kv=0$, as required to retain consistency
with the requirement that $\omega_{i}(0)=N^{2}_{i}$.  

If the high-$q$ behavior of the above integral is regularized introducing
a sharp cutoff at $q=\Lambda$, and using Eq. (\ref{screened_interaction}) 
for all $q < \Lambda$, the resulting integral is found to increase 
linearly with $\Lambda$ as $\Lambda$ is increased. We 
\cite{Grzywacz_Morse} and Beckrich {\it et al.} \cite{Beckrich_07} 
have shown, however, that this UV divergence can be absorbed into 
renormalization of the statistical segment length. After this UV 
divergent part is subtracted, the remaining UV convergent correction, 
which we will denote by $\delta\omega_i^{*}(\kv)$ may be used to 
characterize the deviation from random-walk statistics. More
precisely, $\delta \omega_{i}^{*}(\kv)$ is the deviation of the
intramolecular correlation function $\omega_{i}(\kv)$ from that
of a Gaussian chain with a renormalized statistical segment length 
$b_{\infty,i}$ characteristic of a dense one-component liquid of 
infinitely long chains. The correction $\delta\omega_{i}^{*}(\kv)$ 
is found to be smaller than $\omega_{i}(\kv)$ by a factor proportional 
to $\bar{N}^{-1/2}$, and exhibits a nontrivial dependence on $\kv$.

The fractional deviation of the polymer radius gyration is given
by (appendix \ref{appdx:drg})
\begin{equation}
  \frac{\delta R^2_{g,i}}{R^{2}_{g0,i}} = - \frac{3} { N_i^2 R^{2}_{g0,i} } 
  \lim_{\kv \rightarrow 0} \frac{ \partial \delta \omega_i (\kv) } { \partial (k^2) }. 
  \label{drg_formula}
\end{equation}
where $R^{2}_{g0,i} \equiv N_{i}b_{\infty,i}^{2}/6$ is the 
prediction for a Gaussian chain with statistical segment length
$b_{\infty,i}$. This fractional deviation is proportional to
$\bar{N}^{-1/2}$. 

\subsection{Monodisperse Melt}
\label{sub:SingleChainMelt}

We first consider the case of a monodisperse one-component melt, which 
we obtain by setting $\phi_{2}=0$. The one-loop approximation used 
here reduces in the case of one-component liquids to that investigated 
by Semenov {\it et al.} 
\cite{Semenov_Obukhov_05, Wittmer_07a, Wittmer_07b, Beckrich_07}. 
The only difference in our treatment of this case is our use of a 
numerical integration and renormalization procedure to obtain 
accurate results for monodisperse polymers.

\begin{figure}[htb]\center
\includegraphics[width=0.40\textwidth,height=!]{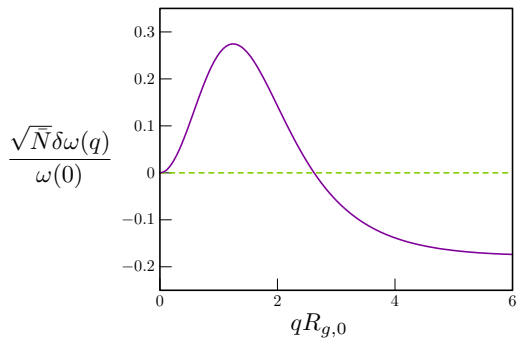}
\caption[One-loop correction to the single chain correlation function.]
{Predicted correction to the single chain correlation function (Debye function).}
\label{domega_theory_k}
\end{figure}

Fig.\ref{domega_theory_k} shows the predicted correction to the single 
chain correlation function in a monodisperse melt.  From the low $\kv$ 
behavior (see appendix \ref{appdx:drg}) we find that the predicted 
fractional change of the radius of gyration is 
\begin{equation}
\frac{\delta R_g^2}{ R_{g0}^2 } = - \frac{ 1.42 } { \sqrt{\bar N} }
\quad. \label{frac_drg}
\end{equation}
We thus predict a contraction of finite chains, relative to a hypothetical 
Gaussian chain with a statistical segment length $b_{\infty}$, in 
agreement with the conclusions of Ref. \cite{Wittmer_07a} [see Eq. (20)].

\subsection{Bidisperse Melt}
\label{sub:SingleChainLongInShort}

We next consider the behavior of a trace concentration of a species 
of chains of length $N$ (component 1) in a matrix of chemically 
homologous shorter chains of length $N_2 = \alpha N$ (component 2).  
This limit is obtained by calculating the limit 
$\phi_{1} \rightarrow 0$ of $\delta\omega_{1}(q)$.  To represent 
chemically similar species, we set $l_1 = l_2$ and $\chi_e = 0$.  
This case reduces to that of a mono-disperse melt when $\alpha = 1$, 
and to a dilute solution of polymer in oligomeric solvent when 
$\alpha \ll 1$. 

\begin{figure}[htb]\center
\includegraphics[width=0.40\textwidth,height=!]{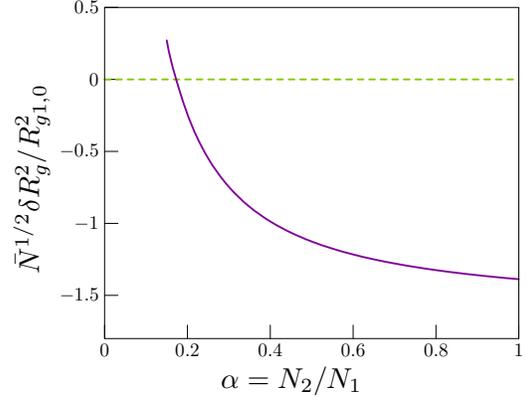}
\caption[Crossover of the single chain size from contraction to swelling.]
{The fractional change of the radius of gyration of an isolated chain of 
length $N$ in a matrix of short chains of length $\alpha N$, rescaled 
with $\sqrt{\bar N}$. } 
\label{alphadrg}
\end{figure}

In the previous subsection, we noted that the size of a single chain in
the pure melt decreases as compared to the unperturbed value. It is 
also well known that a single chain swells in the dilute solutions with 
chemically similar small molecule solvents, which create a good solvent. 
In Fig.\ref{alphadrg}, we show one-loop predictions for the fractional 
change of the radius of gyration of a test chain of length $N$ in a 
matrix of chains of length $\alpha N$. The change is indeed negative 
for $\alpha$ close to 1 and becomes positive for $\alpha$ close 
to 0, passing through zero at $\alpha \simeq 0.17$.

The limit $\alpha \ll 1$ corresponds to a polymer solution in a good
small-molecule solvent. The curve in Fig.\ref{alphadrg} diverges as 
$0.31/\alpha$ near $\alpha = 0$ (appendix \ref{appdx:drg}). In this 
limit, where the matrix chains are point-like, Eq. 
(\ref{screened_interaction}) for the screened interaction between a 
pair of monomers on a longer chain reduces to a constant 
$G_{11}(q) = v / (\alpha N)$. Recall that the definition of a ``monomer" 
in an incompressible coarse-grained theory is simply the length of a 
chain that occupies an arbitrary reference volume $v$.  If, to simplify
notation, the reference volume is taken to be equal to the excluded 
volume $v \alpha N$ of a solvent molecule (i.e., a matrix chain), by 
setting $\alpha N = 1$, we obtain an interaction $G_{11}(q) = v$ 
characterized by an excluded volume parameter equal to the solvent 
volume. Predictions of the theory in this limit are thus identical 
to those obtained in the first order perturbation theory for a single 
chain with a point-like interactions $v\delta(\rv)$.\cite{Yamakawa_book}  
It is straightforward to show that this perturbative result is 
consistent with the prediction \cite{deGennes_book} that the longer 
chain will undergo substantial expansion if $N_{2} \ll \sqrt{N_{1}}$, 
signalling a breakdown of first order perturbation theory.

\subsection{Binary Blends}
\label{sub:SingleChainBinary}

\begin{figure}[htb]\center
\includegraphics[width=0.40\textwidth,height=!]{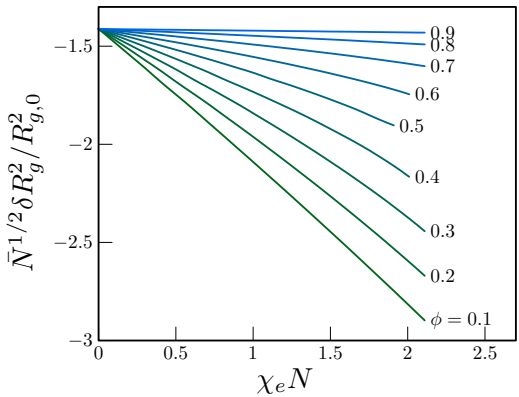}
\caption[One-loop correction to $R_g$.]
{The changes in the radius of gyration of chains at various compositions plotted
as functions of $\chi_e N$ in symmetric binary blends. The chains become more
contracted as $\chi_e N$ increases and the effects is more pronounced for minority
component. In the limit of $\phi = 1$, the change becomes independent of $\chi_e N$;
In the limit of $\phi = 0$, the change is linear in $\chi_e N$, as can be told from
the functional form of $G_i$, Eq.(\ref{intra_corrl}).
}\label{phidrg}
\end{figure}

We now consider binary blends with $\chi \neq 0$. For simplicity, we focus on
symmetric systems, with $N_1 = N_2 = N$ and $b_1 = b_2 = b$.  Fig. \ref{phidrg}
shows the one-loop predictions for the fractional change of the radius of
gyration squared, rescaled by $\sqrt{\bar{N}}$, as function of $\chi_e N$ for
various compositions. Increasing $\chi$ always causes both species in a blend
to contract. The dependence of $R_{g1}$ upon $\chi$ is strongest when species
$1$ is a tracer in a matrix of $2$ ($\phi_{1} \ll 1$).  There is no dependence
of $R_{g1}$ upon $\chi$ in the opposite limit $\phi_{2} \ll 1$ of essentially
pure homopolymer 1.  In the dilute limit, $\phi_{1} \rightarrow 0$, the
one-loop theory predicts a deviation $\delta R_{g1}^{2}$ that is a strictly
linear function of $\chi N$. Furthermore, it can be shown that the slope of the
limiting line is -0.62, which is -2 times that of the coefficient to the
$1/\alpha$ divergence for the bidisperse melt in the limit of small-molecule
good solvent discussed in the previous subsection (appendix \ref{appdx:drg}).

It is interesting to note that at the critical composition, $\phi_{1}
= 1/2$, the deviation induced by increasing $\chi$ always remains less 
than the chain-length dependent deviation from ideal chain statistics 
that is already present in the one-component melt ($\chi=0$). Also
note that the radius of gyration is a smooth, nearly linear function 
of $\chi$ even near the critical point. This reflects the fact that
the long-wavelength composition fluctuations that appear near the
critical point have characteristic wavelengths much larger than the 
radius of gyration, and so couple very weakly to the conformations 
of individual chains.

\section{Comparison to Simulation}
\label{sub:Simulation}
Here, we compare the predictions of the one-loop theory to the results
of lattice Monte-Carlo simulation.  We compare to bond-fluctuation model
(BFM) simulations of symmetric blends at $\phi_{1}=1/2$ Binder, Deutsch, 
and M\"{u}ller, and to BFM simulations of single-chain properties by 
Beckrich {\it et al.} \cite{Beckrich_07}.
The bond fluctuation model is a lattice model on a simple cubic lattice 
in which each monomer occupies a set of 8 sites at the vertices of a 
cube. In all of the simulations discussed here, the monomer concentration
is 1/16 monomer per lattice site, which is half the maximum possible 
concentration in this model. The statistical segment length extrapolated
for the infinitely long chain is 3.244, which has been reported in \cite{Wittmer_07a}.
In simulations of polymer blends, monomers
of types $i$ and $j$ on neighboring sites interact with an energy $u_{ij}$,
with $u_{11}=u_{22}$ and $u_{12} > u_{11}$.  Binder and coworkers have 
carried out semi-grand canonical ensemble simulations using two 
different variants of the rules for what constitute a ``neighbor": 
One in which each monomer can interact only with monomers located at 
the 6 closest allowed sites, and another in which each monomer can 
interact monomers with monomers at any of the 54 closest allowed 
sites. 

\subsection{Determining $\chi_{e}$}
To compare predictions of a coarse-grained theory to simulation data, 
one must somehow establish a relationship between the interaction 
parameters used in the simulations and the phenomenological 
interaction parameter used in both Flory-Huggins theory the one-loop 
theory. In an accompanying paper, we propose a way to do this, which
we use here. The analysis given there refines and justifies a form of 
``modified Flory-Huggins theory" that was originally proposed by 
M\"{u}ller and Binder.\cite{Mueller_Binder_95} 
There, we consider a perturbation theory for a lattice model of the 
type described above, in which free energy of mixing is expanded in 
powers of a perturbation parameter $\alpha = u_{12}-u_{11}$.  
We identify the SCF interaction free energy per monomer $f_{\rm int}$ 
by considering the behavior of the free energy of mixing in the limit 
of infinitely long chains. We show there that $f_{\rm int}$ is given 
to first order in a power series in $\alpha$ by a function 
\begin{equation}
   f_{\rm int}(\epsilon,\phi_{1}) = kT \chi_{e}\phi_{1}(1- \phi_{1})
\end{equation}
in which
\begin{equation}
    \chi_{e} = 
    z^{\infty} [u_{12} - u_{11}]/kT 
    \quad. \label{chie_lattice}
\end{equation}
Here, $z^{\infty}$ is an ``effective coordination number" that 
is obtained by evaluating a value $z(N)$ of the average number of 
inter-molecular nearest neighbors per monomer in a one-component 
reference liquid, with $\alpha=0$, and extrapolating the value to 
the limit of infinitely long chains. The values of $z^{\infty}$ 
used here were reported by M\"{u}ller and Binder: $z_c^\infty = 2.1$ 
for the model with 54 interacting sites, and $z_c^\infty = 0.307$ 
for the model with 6 interacting sites. A more complete  discussion 
of the reasoning underlying this prescription is given elsewhere.

Our use of a first order perturbation theory to estimate $f_{int}$ in 
the one-phase region of the phase diagram is justified by the fact 
that the critical value of $\alpha$ decreases as $1/N$ with increasing 
chain length.  The fractional error in our estimate of $\chi_{e}$ 
arising from the use of a first order expansion is thus expected to 
be of order $1/N$ near the critical point. For $N \gg 1$, this error 
is thus expected to be much smaller than the corrections to $\chi_{a}$ 
predicted by the one loop theory, which are of order $N^{-1/2}$. 
The resulting ${\cal O}(1/N)$ errors in our estimate of the ``bulk" 
interaction parameter $\chi_{e}$ are expected to be comparable in 
importance to the ${\cal O}(1/N)$ corrections that arise from end 
effects.

\subsection{Composition Fluctuations in Blends}
\begin{figure}[htb]
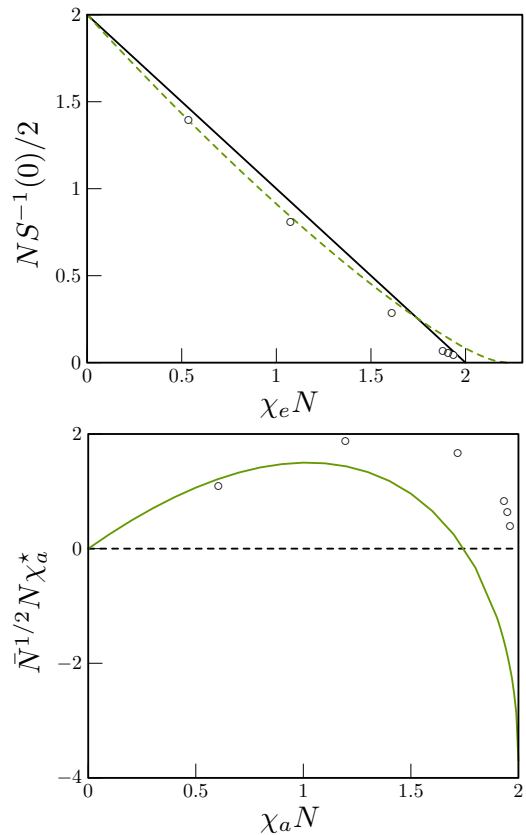
\center
\includegraphics[width=0.40\textwidth,height=!]{Mueller_a.eps}
\includegraphics[width=0.40\textwidth,height=!]{Mueller_b.eps}
\caption[Comparison to M\"{u}ller's lattice data at $\bar N=290$.]
{Comparison of the results of M\"{u}ller's bond-fluctuation model 
lattice simulations $NS^{-1}(0)/2$ (circles) to corresponding 
one-loop 1-loop predictions (dashed line), for $N=64$ and 
$\bar N = 290$, for a half-filled lattice \cite{Mueller_Pablo_06}. 
The calculation of values of $\chi_{e}$ for simulation results is 
discussed in the text.} 
\label{Mueller_Binder_S0}
\end{figure}

Fig. \ref{Mueller_Binder_S0} shows the results of M\"{u}ller for 
$NS^{-1}(0)/2$ plotted vs. $\chi_{e}N$ for bond-fluctuation model 
simulations of chains of length $N=64$ \cite{Mueller_Pablo_06}.
These simulation used a variant of the BFM in which interactions 
extend over 54 neighboring sites, and in which $u_{11}=-u_{12}$. 
Values of $S^{-1}(0)$ were extracted from the semi-grand equation 
of state. We have calculated the SCF effective $\chi_{e}$ parameter 
for each simulation from Eq.  (\ref{chie_lattice}) using the 
value of $z^{\infty}= 2.1$ reported by Mueller and Binder. 
\cite{Mueller_Binder_95}

In the upper panel of Fig. \ref{Mueller_Binder_S0} the solid line 
is the RPA prediction and the dashed line is the one-loop prediction
obtained using the appropriate value of $\bar{N} = 290$ (The
value was reported as 240, which were computed from the statistical
segment length of $N=64$ chains). The lower
shows a comparison of simulation results for the difference 
$\chi^{*} N = (\chi_{a}-\chi_{e})N$ multiplied by $\sqrt{\bar N}$ 
to the universal curve shown in Fig. \ref{dchiN}. We see that 
the self-consistent one-loop theory gives an extremely accurate 
description of deviations from SCFT far from the critical point, 
for $\chi_{a}N \alt 1$, but appears to overestimate the stabilizing 
effect of critical fluctuations. For this system, the suppression 
of $\chi_{a}$ by long-wavelength fluctuations near the critical 
point is almost exactly cancelled by the enhancement of $\chi_{a}$ 
at smaller values of $\chi_{e}$, giving a critical value of
$\chi_{e}N$ that is closer to the SCFT value than to the value 
predicted by the one-loop theory. 

\begin{figure}[htb]\center
\includegraphics[width=0.40\textwidth,height=!]{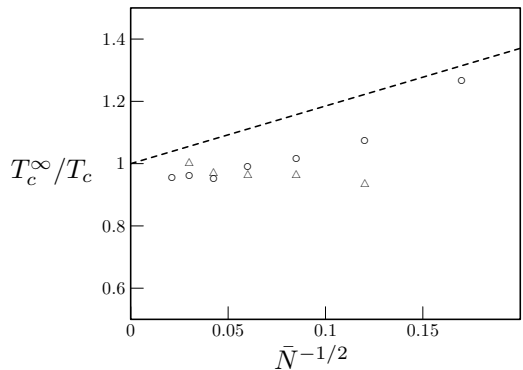}
\caption[Chain length dependence of the shift in the critical temperature.]
{Chain length dependence of the shift in the critical temperature.
Comparison between 1-loop theory prediction and Deutsch and Binder's 
lattice simulation \cite{Deutsch_Binder_93}. Simulation results are
shown for variants of the BFM in which interactions extend over 6
nearest neighbors (circles) and 54 neighbors (triangles). }
\label{Deutsch_Binder_Tc}
\end{figure}

Fig. \ref{Deutsch_Binder_Tc} compares the predictions of the 
one-loop theory for the ration $T_{c}^{\infty}/T_{c}(N)$ for 
different values of $N$ to the results of simulations by 
Deutsch and Binder \cite{Deutsch_Binder_93}. Here, 
$T_{c}^{\infty} \equiv \alpha z^{\infty}N/(2k)$ is the 
critical temperature that is predicted by SCF if we use 
Eq. (\ref{chie_lattice}) for $\chi_{e}$. The critical temperatures 
reported by Deutsch and Binder were identified for each chain 
length by a finite-size scaling analysis of simulations with 
different simulation cell sizes. Simulation results are shown 
for variants of the BFM model in which the neighbor interactions 
extend over 6 neighbors (circles) and over 54 neighbors (triangles),
For both versions of the models, $T_{c}^{\infty}/T_{c}(N)$ 
extrapolates to unity as $N \rightarrow \infty$. This confirms 
that first order perturbation theory prediction for $T_{c}$ is 
indeed asymptotically exact in the limit $N \rightarrow \infty$.

For both versions of the BFM, however, the measured deviations of 
$T_{c}^{\infty}/T_{c}(N)$ are significantly smaller than those 
predicted by the one-loop theory. This quantitative discrepancy
must be, in part, a result of the inadequacy of the one-loop 
theory as a theory of critical phenomenon, and the need for a 
renormalization group approach near the critical point.  This 
data suggests that, if it is indeed possible to express universal 
corrections to $T_{c}^{\infty}/T_{c}(N)$ that arise from critical 
fluctuations as a power series in $1/\sqrt{N}$, the coefficient 
of any ${\cal O}(N^{-1/2})$ correction must be much smaller than 
predicted by the one-loop theory. 

It is important to also note that the slight deviations of 
$T_{c}^{\infty}/T_{c}(N)$ from unity in Fig. \ref{Deutsch_Binder_Tc} 
are not the same for the two variants of the BFM, and appear to 
of opposite sign.
This model-dependence cannot be so easily explained as a failure 
of the one-loop approximation near the critical point, since we 
expect that a more sophisticated treatment of critical phenomena 
in a simple coarse-grained model would also predict a universal 
dependence of $T_{c}^{\infty}/T_{c}(N)$ on the dimensionless
parameter $\bar{N}$. The observed non-universality could, 
however, be the result of corrections arising from contribution 
to $f_{\rm int}$ that are second order in $\alpha$, which are 
neglected in our estimate of $\chi_{e}$, and/or the result of 
end effects. Both of these effects are neglected in our analysis, 
and both are expected to yield model dependent corrections to 
$T_{c}/T_{c}^{\infty}$ that are proportional to $1/N$. It 
appears to us that the dominant contribution to the deviation 
$T_{c}^{\infty}/T_{c}(N) - 1$ for this data may well be a 
nonuniversal correction proportional to $1/N$ (which appears to 
provide a better description of the $N$ dependence, as originally 
suggested by Deutsch and Binder) rather than a universal correction 
proportional to $1/\sqrt{\bar{N}}$.

\subsection{Single-Chain Correlations}
\begin{figure}[htb]\center
\includegraphics[width=0.40\textwidth,height=!]{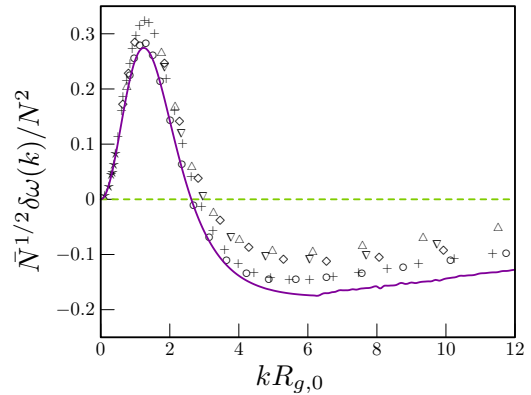}
\caption[One-loop correction to the single chain correlation function.]
{One-loop correction to the single chain correlation function (Debye function).
The data points are gleaned from Fig.6 of Ref.\cite{Beckrich_07}. 
Different symbols are lattice simulation ($\rho = 1/16$, $b \simeq 3.24$) 
results for chains with different lengths: $+$ ($N$ = 2048),
$\circ$ (1024), $\diamond$ (512), $\triangledown$ (256),
\scalebox{.7}{\mbox{$\triangle$}} (128). Near $k=0$ the data points 
for various chain lengths overlap and can not be distinguished from 
the original figure, so are symbolized as $\star$. }
\label{domegak}
\end{figure}

In Fig.\ref{domegak} we compare one-loop predictions and simulation
results of Beckrich {\it et al.} \cite{Beckrich_07} for the deviation 
$\delta\omega(k)$ of the correlation function for a monodisperse melt 
from the Debye function.  Simulation results for 5 different chain 
lengths were taken from Ref. \cite{Beckrich_07}. 
As noted by Beckrich {\it et al.}, the collapse of data for from 
different chain lengths when $\sqrt{\bar{N}}\delta \omega(q)$ is
plotted vs. $kR_{g0}$ already shows that the deviation is proportional 
to $1/\sqrt{\bar{N}}$, as predicted by the one-loop theory. 
Simulations and predictions agree quite well, though there do appear
to be some small systematic discrepancies in the high $\kv$ regime. 
The agreement is somewhat better than that obtained by the Strasbourg
group using a more approximate treatment of a monodisperse melt. 
A similar level of agreement was obtained previously by that group
in comparisons of the one-loop theory to simulations of polydisperse 
equilibrium polymers, for which they obtained analytic predictions 
for $\delta\omega(q)$. 
\cite{Semenov_Obukhov_05,Wittmer_07a, Wittmer_07b, Beckrich_07,
Shirvanyants_Rubinstein_08}.

\begin{figure}[htb]\center
\includegraphics[width=0.40\textwidth,height=!]{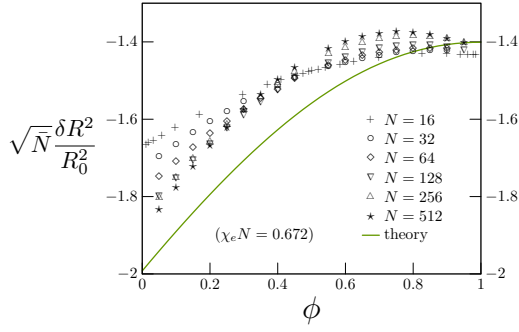}
\caption[Shrinkage of the mean squared end-to-end distance of chains in blends.]
{Composition dependence of the fractional change of the mean squared end-to-end
distance for homopolymer chains in a binary blends for systems with $\chi_{e} N=
0.672$ and variable composition. Simulation data were adapted from
M\"{u}ller's data for $\langle R^2 \rangle$ \cite{mueller_98}. Solid
line is the one-loop prediction.}
\label{compR}
\end{figure}

M\"{u}ller has examined the dimension of homopolymer chains of various 
lengths in binary blends with varying compositions and interactions
\cite{mueller_98}, using Monte Carlo simulations of the BFM with the
same parameters as the one used to study composition fluctuations 
\cite{Mueller_Pablo_06}. In Fig.\ref{compR} we re-plotted 
$\sqrt{\bar{N}}$ times his results for the fractional deviation of the 
mean-squared end-to-end vector $R^2$ for component $A$ over a range of 
values of $\phi = \phi_{A}$ for a fixed value $\alpha N/k_BT = 0.32$, 
corresponding to $\chi_{e} N = 0.672$, for 6 different chain lengths, 
adapted from Fig.5b in \cite{mueller_98}. Here $\delta R^{2}$ represents
the deviation of $R^{2}$ in a mixture from the ideal Gaussian chain value
calculated using the statistical segment length of a sufficiently long
chain. The original results were presented as the difference between
the value in the blend and that in a melt of chains of the same length. 
We have switched the normalization by assuming that the radius of gyration
measured in the melt is smaller compared to the ideal value by the amount
$1.42/\sqrt{\bar N}$.
The fact that results for different chain lengths 
collapse confirms by itself that these deviations are proportional to 
$1/\sqrt{N}$. The method of obtaining the theoretical curve in Fig.\ref{compR}
is sketched in appendix \ref{appdx:dr}. The qualitative trend of the
simulation results were relatively well described with the theory.

\section{Thermodynamic Integration of the RPA}
\label{sec:ThermoInt}
The one-loop approximation has most often been presented within 
the context of the Edwards formalism, as a Gaussian approximation 
for the functional integral representation for the partition 
function. Here, we discuss an alternative derivation that makes 
no reference to the Edwards formalism, in which the one-loop free 
energy is obtained by thermodynamic integration with respect to 
the strength of either the pair potential or the bare interaction
parameter, while using the RPA approximation for $S_{ij}(q)$. 
The derivation is closely analogous to the usual derivation 
of the Debye-H\"{u}ckel approximation for excess free energy 
of an electrolyte solutions, which is obtained by combining 
an expression for the free energy as an integral with respect 
to a "charging" parameter with the Debye-H\"{u}ckel linear 
self-consistent field approximation for the correlation function 
(see, {\it e.g.}, McQuarrie \cite{McQuarrie_book}, pgs. 328-339).

\subsection{Variation of the Pair Potential}
\label{app:RelationRPA}
Our analysis starts with the following exact expression for the
dependence of the free energy $F = -\ln Z$ upon the pair potential
$U_{ij}(\rv-\rv')$ in any model with a pair potential: The perturbation 
$\delta F$ in the free energy induced by an infinitesimal perturbation 
$\delta U_{ij}(\rv-\rv')$ in the pair potential is given by an integral
\begin{equation}
   \delta F = \frac{1}{2} \int d\rv \int d\rv' \;
   \langle c_{i}(\rv)c_{j}(\rv') \rangle
   \; \delta U_{ij}(\rv-\rv')
\end{equation}
The corresponding perturbation of the correlation free energy density
$f_{\rm corr}$ may expressed as an integral 
\begin{equation}
    \delta f_{\rm corr} = \frac{1}{2} \int d\rv \;
    S_{ij}(\rv) \; \delta U_{ij}(\rv) 
\end{equation}
or as a Fourier integral
\begin{equation}
    \delta f_{\rm corr} = \frac{1}{2}
    \int_{\qv} S_{ij}(\qv) \; \delta U_{ij}(\qv) 
    \quad, \label{FcorrVariation}
\end{equation}
where $S_{ij}(\rv) \equiv \langle c_{i}(\rv) c_{j}(0) \rangle -  
\langle c_{i}\rangle \langle c_{j} \rangle$,

By explicitly evaluating the corresponding variation of 
Eq. (\ref{dfOneLoopBare0}) for the one-loop correlation free energy 
for a compressible liquid, we find that, in this approximation,
\begin{equation}
    \delta f_{\rm corr} = \frac{1}{2}\int_{\qv} 
    S^{\rm RPA}_{ij}(\qv) 
    \delta U_{ij}(\qv) 
\end{equation}
where
\begin{equation}
   S^{\rm RPA}_{ij}(\qv) =
   [ \Ssh^{-1}(\qv) + \Uh(\qv) ]^{-1}_{ij}
\end{equation}
is the RPA (or mean-field) approximation for $S_{ij}(\qv)$ 
in a compressible liquid.
We thereby find that the one-loop approximation for the free 
energy density is thermodynamically consistent with an RPA 
approximation for $S_{ij}(\qv)$.  

It follows that the one-loop free energy may be obtained by
a thermodynamic integration similar to that used to obtain
the free energy in the Debye-H\"{u}ckel theory of electrolytes:
Consider a sequence of systems with a reduced pair potential
$\lambda U_{ij}$, with a charging parameter $\lambda$ in 
the range $0 < \lambda < 1$.  The one-loop correlation 
energy (which vanishes in the $\lambda=0$ ideal gas state)
may be expressed as an integral
\begin{equation}
   f_{\rm corr} = 
   \frac{1}{2} \int_{0}^{1} d\lambda \int_{\qv} 
   U_{ij}(\qv) S_{ij}^{RPA}(\qv;\lambda) 
\end{equation}
in which $S_{ij}^{RPA}(\qv;\lambda)$ is the RPA correlation
function for a compressible mixture with a reduced pair 
potential $\lambda U_{ij}$.  It is straightforward to confirm 
that Eq. (\ref{dfOneLoopBare0}) is recovered by explicitly 
completing the integral with respect to $\lambda$.

\subsection{Variation of $\chi_{0}$}
For many purposes, it is useful to consider the simple special 
case of a symmetric mixture of structurally identical chains,
with $b_{1}=b_{2}$ and $N_{1}=N_{2}$, with a pair potential of
the form given in Eq. (\ref{Uijmatrix}).  In this case, the limit 
$\chi_{0}=0$ yields a true ideal mixture, for which $f_{\rm corr}$ 
is independent of composition. In a nearly-incompressible 
symmetric mixture, with $Bv \gg 1$, the composition 
dependence of $f_{\rm corr}$ may thus be obtained by starting 
from a nearly incompressible reference state with $\chi_{0}=0$, 
and calculating how $f_{corr}$ varies as $\chi_{0}$ is increased. 
As a special case of Eq. (\ref{FcorrVariation}), it is 
straightforward to show that
\begin{equation}
   \frac{\partial f_{\rm corr}}{\partial \chi_{0} }  = 
   v \int d\rv \; S_{12}(\rv)
   \delta_{\Lambda}(\rv) \quad.
   \label{dfcorrdchi}
\end{equation}
If we take the incompressible limit, for which 
$S_{12}(\rv) = -S(\rv)$, and the limit 
$\Lambda \rightarrow \infty$ of point-like interactions, Eq. 
(\ref{dfcorrdchi0}) reduces to
\begin{equation}
   \lim_{\Lambda \rightarrow \infty} 
   \frac{\partial f_{\rm corr}}{\partial \chi_{0} } 
   = - v \lim_{\rv \rightarrow 0} S(\rv)
   \quad, \label{dfcorrdchi0Point}
\end{equation}
or
\begin{equation}
   \frac{\partial f_{\rm corr}}{\partial \chi_{0} } = 
   - v\int_{\qv} S(q)
   \quad. \label{dfcorrdchi0}
\end{equation}
in Fourier space. 

The one-loop theory may be obtained by using the incompressible
RPA approximation for $S(q)$ in Eq. (\ref{dfcorrdchi0}).  In a 
symmetric blend, the incompressible RPA yields
\begin{equation}
    S^{RPA}(q;\chi) = 
    \frac{\phi_{1}\phi_{2}}{\Ssh^{-1}(q) - 2v \chi \phi_{1}\phi_{2} }
    \quad. \label{S_RPA_symmetric}
\end{equation}
An idealized IRPA model of Gaussian chains with point-like 
interactions, however, yields a UV divergent integral in
Eq. (\ref{dfcorrdchi0}). This may be regularized by restricting 
the integral to $q < \Lambda$. The regularized (but unrenormalized)
one-loop free is then given, to within a composition independent 
constant, by an integral
\begin{equation}
   f_{\rm corr} = - v  \chi_{0}
   \int_{0}^{1}d\lambda \int_{\qv} 
    S^{RPA}(q;\lambda\chi_{0})
   \quad, \label{fcorr_int}
\end{equation}
where the Fourier integral is constrained to $|\qv| < \Lambda$.

\section{Correlation Hole Effects}
\label{sec:CorrHole}
Several features of the one-loop approximation for $f_{\rm corr}$ 
may be understood by relating them to corresponding features of the 
incompressible RPA description of the so-called correlation hole. 
To show this, it is useful to consider the behavior of the 
derivative $\partial f_{\rm corr}/\partial \chi_{0}$ in the simple
case of an ideal statistical mixture, which is obtained in our model 
by setting $b_{1}=b_{2}$, $N_{1}=N_{2}$, and $\chi_{0}=0$. In this 
limit, Eq. (\ref{S_RPA_symmetric}) for $S(q)$ reduces to 
\begin{equation}
   S(q) = \phi_{1}\phi_{2}\Ssh(q)
   \quad. \label{S_chi=0}
\end{equation}
This expression can also be obtained more directly by the following 
physical arguments:
Let $S(r) = \langle \delta c(r) \delta c(0) \rangle$ be the 
overall correlation function in a one-component reference liquid
of either pure $1$ or pure $2$, and let $H(r)$ and $\Ssh(r)$ be 
the inter-molecular and intramolecular correlation functions, so 
that $S(r) = \Ssh(r) + H(r)$.  An ideal mixture may be created 
by randomly labelling a fraction $\phi_{1}$ of the chains in such
a one-component liquid as species ``$1$", and the remainder as 
species ``$2$". In the resulting ideal mixture, the probability 
that a particular pair of monomers on different chains will be 
on chains labelled $i$ and $j$ is simply $\phi_{i} \phi_{j}$.
It follows that the inter-molecular correlation function $H_{ij}(r)$ 
in such a mixture is simply $H_{ij}(r) = \phi_{i}\phi_{j}H(r)$.
Similarly, because the probability that two monomers on the
same chain belong to a chain of type $i$ is $\phi_{i}$, the
intra-chain correlation function $\Ssh_{ij}(r)$ in such a 
mixture is $\Ssh_{ij}(r) = \delta_{ij}\phi_{i}\Ssh(r)$. Thus 
far, everything we have said is exact. 

In the idealized limit of a nominally incompressible liquid,
however, we assume that $S(r)=0$, and thus that $\Ssh(r) + H(r) 
= 0$.  In an incompressible liquid, the inter-chain correlation 
function $H(r)$ must thus have a ``correlation hole", which is a 
mirror image of the intra-chain correlation function $\Ssh(r)$. 
This implies that, in an incompressible statistical mixture, 
$H_{ij}(q) = -\phi_{i}\phi_{j}\Ssh(q)$. Using this assumption, 
it is straightforward to show that
$S_{12}(q) = - \phi_{1}\phi_{2}\Ssh(q)$ and that
$S_{11}(q) = + \phi_{1}\phi_{2}\Ssh(q)$ in any incompressible
ideal mixture. 

\subsection{UV divergence (Unrenormalized Theory)}
The UV divergence of the correlation correction to $\chi_{e}$ 
in Eq. (\ref{chie_sym}) is the result of a pathological feature 
that arises from a combination of the assumption of rigorous
incompressibility, at all wavelengths, in a model of continuous
Gaussian chains.  In the limit $\chi_{0}=0$ and 
$\Lambda \rightarrow \infty$, Eqs. (\ref{dfcorrdchi0}) and 
(\ref{S_chi=0}) may be combined to show that
\begin{equation}
   \frac{\partial f_{\rm corr}}{\partial \chi_{0} } = 
   v \phi_{1}\phi_{2} H(r=0)
   = - v \phi_{1}\phi_{2} \Ssh(r=0)
   \quad,
\end{equation}
In the limit $\Lambda \rightarrow 0$, the predicted free energy 
is sensitive to the value of $H(r)$ only at $r=0$ because the
range of the pair interaction $U_{ij}$ vanishes in this limit.
It is well known, however, that the intramolecular correlation 
function $\Ssh(r)$ for a continuous Gaussian thread diverges as 
$\Ssh(r) \propto 1/r$ as $r \rightarrow 0$. Recall the origin
of this divergence: The number of monomers within a region of 
volume $r^{3}$ near a test monomer is of order the number of 
monomers $g$ in a segment of chain of size $r = \sqrt{g}b$, 
giving an average concentration $g/r^{3} \sim 1/(rb^{2})$ within
a region of size $r$, or $\Omega(r) \propto 1/r$. When combined 
with an assumption of rigorous incompressibility, however, the 
divergence in $\Ssh(r)$ yields an infinitely deep correlation 
hole in $H(r)$ at $r=0$ in the one-component reference liquid. 
Taken literally, the incompressible RPA would yield negative 
values of $\langle c(r) c(0) \rangle$ in a region around $r=0$. 
This leads to the nonsensical prediction of a divergent reduction 
in $H(r=0)$, and thus to a divergent negative correction to 
$\chi_e$. 

The UV divergence obtained from the IRPA that underlies the 
one-loop theory could be avoided by the use of any model 
that avoids the use of any of: i) a point-like interaction, 
ii) a continuous Gaussian thread model, or iii) rigorous 
incompressibility. For example, in the thread limit of the 
PRISM theory \cite{Schweizer1990}, one retains a model of
continuous Gaussian chains, and a point-like effective
interaction (i.e., direct correlation function), but avoids
complete nonsense by allowing for some compressibility at
short wavelengths.

\subsection{$N$-dependence (Renormalized Theory)}
In the renormalized one-loop theory, the UV divergent part 
of the correlation correction to $\chi_{a}$ is absorbed into 
the definition of $\chi_{e}(\Lambda)$. Once the divergence 
is removed, the renormalized theory can make concrete predictions 
about the difference $\chi^{*} = \chi_{a}-\chi_{e}$.
Because the one-loop expression for $\chi^{*}$ is proportional
to $\bar{N}^{-1/2}$, $\chi_{e}$ may be interpreted as the 
limit of $\chi_{a}$ extrapolated to the limit $N \rightarrow
\infty$.  We now show that the value of $\chi^{*}$ is 
sensitive to the slight difference between the depth of the 
correlation hole in a system of finite chains and that in a 
hypothetical system of infinite chains.

To show this, consider the derivative of Eq. (\ref{dfOneLoop})
for $f^{*}(\chi_{a})$ with respect to $\chi_{a}$, evaluated 
(for simplicity) in a reference state with $\chi_{a}=\chi_{0}=0$. 
A straightforward differentiation yields
\begin{equation}
   \lim_{\chi_a \rightarrow 0} 
   \frac{\partial f^{*}}{\partial \chi_{a} } = 
   - v \phi_{1}\phi_{2} 
   \int_{\qv}^{*} \Omega(q)
   \quad. \label{dfstardchia}
\end{equation}
The only differences between Eq. (\ref{dfstardchia}) and the 
limit $\chi_{0}=0$ of Eq. (\ref{dfcorrdchi0}) for 
$\partial f_{\rm corr}/\partial \chi_{0}$ are: 
i) the use of a renormalized integral (denoted by an asterisk) 
to remove the UV divergence,
ii) the use of $\chi_a$ rather than $\chi_{0}$ as the input
parameter in a Hartree approximation. 

The UV divergent part of the integral in Eq. (\ref{dfstardchia})
is given by a difference
\begin{equation}
   \lim_{\chi_{a} \rightarrow 0} 
   \frac{\partial f}{\partial \chi_{a} } = 
   - v^{2} \phi_{1}\phi_{2}
   \int_{\qv} \; [ \Omega(q) - \Omega^{\Lambda}(q) ]
   \quad, \label{dfchistardchia}
\end{equation}
in which $\Omega^{\Lambda}(q)$ is an asymptotic approximation 
for $\Omega(q)$ in the limit $q \gg R^{-1}$. This quantity is
given by
\begin{equation}
   \Omega^{\Lambda}(q) \equiv 12/(v q^{2}b^{2})
   \quad .
\end{equation}
Let $\Omega(q;N)$ denote the intramolecular correlation 
function for a single-component melt of chains of length $N$. 
Because $\Omega^{\Lambda}(q)$ is an asymptotic 
expansion of $\Omega(q;N)$ in the limit $qR \gg 1$, it may be 
interpreted equally well as either:
\begin{itemize}
\item[i)] The high-$q$ behavior of $\Ssh(q;N)$ for chains of 
fixed length $N$, or 
\item[ii)] The asymptotic behavior of $\Ssh(q;N)$ at fixed $q$ 
in the limit $N \rightarrow \infty$. 
\end{itemize}
By adopting the latter interpretation, and inverting the Fourier
transform, Eq. (\ref{dfchistardchia}), may thus be expressed as 
a difference
\begin{equation}
   \lim_{\chi_a \rightarrow 0} 
   \frac{\partial f^{*}}{\partial \chi_{a} } 
   = -v \phi_{1} \phi_{2} \lim_{\rv \rightarrow 0}
   [ \Ssh(\rv;N) - \Ssh(\rv;\infty) ]
   \label{dfchistar_dchia_r}
\end{equation}
or, equivalently,
\begin{equation}
   \lim_{\chi_a \rightarrow 0} 
   \frac{\partial \chi^{*}}{\partial \chi_{a} } 
   = v^{2} \lim_{\rv \rightarrow 0}
   [ H(\rv;N) - H(\rv;\infty) ]
   \label{dfchistar_dchia_r_H}
\end{equation}
where $H(\rv;N)$ denotes the value of the inter-molecular correlation 
function $H(\rv)$ in a one-component melt of chains of length $N$.  

The renormalization procedure that was originally introduced remove 
the UV divergence is thus seen to be equivalent to the subtraction 
of the $N \rightarrow \infty$ limit. 
The resulting difference relates $\chi^{*} = \chi_a - \chi_e$ to the
difference in the depth of the correlation hole at $\rv=0$ in a system 
of finite chains and that in a hypothetical system of infinite chains. 
The renormalized coarse-grained theory correctly predicts the behavior 
for $\chi^{*}$ for small values of $\chi_a$ because it correctly 
describes this subtle difference between finite and infinite chains, 
despite the fact that it predicts a UV divergence $H(\rv;N)$ itself.


\section{Conclusions}
\label{sec:Conclusions}

We have use the renormalized one-loop theory to analyze corrections to 
the RPA description of composition fluctuations in binary blends, and 
to the random walk model of chain conformations. 

Our treatment of long-wavelength composition fluctuations is closely
related to an earlier treatment by Wang, who focused primarily on behavior 
near the spinodal.\cite{Wang_02} Our predictions can be summarized by 
considering the difference $\chi^{*} \equiv \chi_{a} - \chi_{e}$ between 
an apparent interaction parameter $\chi_{a}$ that would be inferred from 
the magnitude $S(q=0)$ of long-wavelength composition fluctuations and a 
parameter $\chi_{e}$ that we identify as the SCFT interaction parameter, 
which is the limit of $\chi_{a}$ as $N \rightarrow \infty$. 

We show that the deviation from SCFT can be understood as the result of 
two competing physical effects: In the limit $\chi_{e}N \ll 1$ of a 
nearly ideal mixture, $\chi_{a}$ is greater than $\chi_{e}$ because of 
a simple packing effect: Monomers on a finite chain are slightly less 
strongly shielded from contact with monomers of other chains than are 
monomers on an infinite chain, and are therefore more exposed to 
unfavorable interactions with chains of the other species in a binary 
blend. We discuss this effect from another point of view in an 
accompanying paper.  Closer to the spinodal, however, the build up of 
long-wavelength correlations causes the difference $\chi_{a}-\chi_{e}$ 
to decrease with increasing $\chi_{e}$.  Near the critical point, both 
effects yields contributions to $\chi_{a}$ that are proportional to 
$\bar{N}^{-1/2}$, and comparable in magnitude. 
An accurate prediction of the magnitude and sign of the absolute 
shift in the critical value of $\chi_{e}N$ thus requires an accurate 
description of both of these competing effects. A simple renormalized
Hartree theory predicts an overall enhancement in the critical value 
$(N\chi_{e})_{c}$ (or a depression in $T_{c}$) for a symmetric blend 
by an amount proportional to $\bar{N}^{-1/2}$, as first suggested by 
Holyst and Vilgis. \cite{Holyst_Vilgis_93, Holyst_Vilgis_94}

Comparison to the data from lattice Monte Carlo simulations of symmetric
blends at the critical composition indicate that the one-loop theory 
gives an excellent description of the packing effect that dominates
corrections to the RPA for $\chi_{e}N \alt 1$, but a mediocre description 
of correlation effects closer to the critical point. It appears that 
the self-consistent one-loop (or Hartree) theory gives a reasonable 
semi-quantitative description of deviations from SCFT, but tends to 
overestimate the magnitude of the depression in $\chi_{a}$ near the 
critical point, and thus to predict too high a value for $(\chi_{e}N)_{c}$. 

Predictions for corrections to random-walk statistics were also compared
to the available simulation data. Predictions for deviations of the
intramolecular correlation function from the Debye function for chains in 
a monodisperse single-component melt agree very well with the simulation 
results of Beckrich {\it et al.}, \cite{Beckrich_07} and deviate only 
slightly from a more approximate theoretical treatment of monodisperse 
polymers by the same group. Predictions of the composition dependence 
of the radius of gyration in a binary blend with $\chi_{e} \neq 0$ show 
a similar level of quantitative agreement with the results of lattice 
Monte Carlo simulations. 


\appendix
\section{Radius of Gyration}
\label{appdx:drg}

The $k$-dependence of the correction to the intramolecular correlation function
of a homopolymer, Eq.(\ref{intra_corrl}), comes from that of
$\psi_i^{(4)}(\kv,-\kv,\qv,-\qv)$. A straightforward Taylor expansion of
$\psi_i^{(4)}$ shows that in the low $\kv$ regime, to second order in an
expansion in powers of $|\kv|$,
\begin{equation}
   \psi_i^{(4)} \simeq N_i^4 f(q^2 R^2_{g0,i}) k^2R^2_{g0,i} \cos^2\theta
\end{equation}
where $\theta$ is the angle between vectors $\kv$ and $\qv$, and
\begin{eqnarray}
   f(x) &\equiv& \frac{2}{3}
   \left( \frac{1}{x^2} - \frac{24}{x^4} - \frac{96}{x^5} - \frac{120}{x^6} \right)
   \ee^{-x} \nonumber\\
   & + &
   \frac{8}{3} \left(
   \frac{1}{x^3} - \frac{3}{x^4} - \frac{6}{x^5} + \frac{30}{x^6} \right)
\end{eqnarray}
This expansion has been obtained previously in the studies of chain sizes 
in semi-dilute solutions\cite{Ohta_Nakanishi_83} and in 
melts. \cite{aksimentiev_holyst_98} 
Using the fact that $\omega(\kv) \simeq N_i^2(1 - k^2 R_{g,i}^2
/3)$ in the low-$\kv$ region and Eq.(\ref{intra_corrl}), one finds:
\begin{equation}
   \frac{\delta R^2_{g,i}}{R^{2}_{g0,i}}
   = \frac{3 N_i^2}{2} \int^{*}_{\qv} f(q^2 R^2_{g0,i}) \cos^2\theta G_{ii}(\qv)\quad .
   \label{dRgSquared}
\end{equation}
This is the general expression used to calculate the fractional change of the
radius of gyration.

Consider the variation of $R_g^2$ for a tracer amount of polymer species $1$ 
in a matrix of chains of species $2$.  In the relevant limit 
$\phi_1 \rightarrow 0$, the screened interaction $G_{11}$ reduces to:
\begin{equation}
   G_{11}(q) = \frac{v}{N_2} 
   \left( \frac{1}{\Debye(q^{2}R_{g0,2})} - 2\chi_e N_2 \right),
\end{equation}
where $\Debye(x) = 2(e^{-x}-1+x)/x^{2}$ is the Debye function and
$R_{g0,2} = N_{2}b^{2}_{2}/6$. Substituting this expression into 
Eq.(\ref{dRgSquared}), and hereafter taking $b_{1}=b_{2}=b$, we 
find:
\begin{equation}
    \sqrt{\bar{N}_1} \frac{\delta R^2_{g,1}}{R^2_{g0,1}}
    = \frac{F(\alpha)}{\alpha} - 2 F(0) \cdot \chi_e N_1
\end{equation}
where
\begin{equation}
   F(\alpha) \equiv \frac{6^{3/2}}{4\pi^2}
   \int^*\! dy \frac{f(y^2)y^2}{\Debye(\alpha y^2)}
\end{equation}
where $\alpha \equiv N_2/N_1$, and
where $y \equiv q^2 R^2_{g0,1}$.  
$F(\alpha)$ is convergent when $\alpha = 0$ and is UV-divergent for any
non-vanishing $\alpha$. In the latter case, a constant $4/3$ has to be
subtracted from the integrand to account for the renormalization of the
statistical segment length.

Three special cases of this result are noted in the main text: The fractional 
shrinkage of $R_g^2$ in a homopolymer melt, which is discussed in subsection 
\ref{sub:SingleChainMelt}, is obtained by setting $\chi_e = 0$ and $\alpha=1$. 
This yields $\sqrt{\bar{N}} \delta R_g^2 / R_{g,0} = F(1) = -1.42$.  In 
section \ref{sub:SingleChainLongInShort}, the curve shown in
Fig. \ref{alphadrg} is given by $F(\alpha)/\alpha$. Its asymptotic behavior
in the limit of a small molecule solvent, $\alpha \rightarrow 0$, is given 
by $ F(0)/\alpha = 0.31 / \alpha$.  In section \ref{sub:SingleChainBinary}, 
the $\chi_e N$ dependence of the fractional change of $R^2_g$ in the limit 
of $\phi_1 \rightarrow 0$ is given by 
$\partial (\sqrt{\bar{N}} dR^2_g / R^2_{g,0}) / \partial (\chi_e N) =  - 2
F(0) = -0.62$.

\section{Mean-squared End-to-End Distance}
\label{appdx:dr}

We calculate the one-loop correction to the mean-squared end-to-end 
distance $R^2$ using the method described in \cite{Doi_Edwards} (appendix 
III of chapter 5). There, the same quantity was calculated for a finite 
segment of an infinitely long chain in a semi-dilute solution, using a 
Lorentzian approximation for the screened interaction. The analogous 
result for a chain of species $1$ in an incompressible binary blend is 
obtained by using the appropriate screened interaction $G_{ij}$. The 
final result for a symmetric blend ($N_{1}=N_{2}$ and $b_{1}=b_{2}$) is
\begin{eqnarray}
   \frac{\delta R^2_{1}}{R_0^2}
   = \frac{\sqrt{24}}{\pi^2\sqrt{\bar{N}}} \int^*\!\!\! dy 
     \frac { h(y^2) y^4 } {\Debye(y^2)}
     \frac{1 - 2 \phi_2 \chi_e N \Debye(y^2)}{1 - 2\phi_1\phi_2 \chi_e N
     \Debye(y^2)}
\end{eqnarray}
where $y \equiv q^2 R_{g0}^{2}$, and
\begin{equation}
   h(x) \equiv
   \left( \frac{1}{x^2} + \frac{4}{x^3} + \frac{6}{x^4} \right) \ee^{-x}
   + \frac{2}{x^3} - \frac{6}{x^4} \quad.
\end{equation}
This expression reduces to Eq. (15) of Ref. \cite{Wittmer_07a} for a 
one-component melt ($\phi_2 = 0$).

The integral is divergent in the high-$q$ region, and the dangerous term 
in the integrand is ``1''. This is precisely the divergent term 
identified by Edwards, which he identified with as a renormalization of 
the statistical segment length. \cite{Doi_Edwards}. We subtract this 
term from the integrand in order to obtain a UV convergent prediction 
for $\delta R^{2}$, which is the deviation of $R^{2}$ from the value 
$R_{0}^{2}(N)$  predicted by a Gaussian model with a statistical segment 
length equal to that obtained for infinitely long chains. For a pure 
homopolymer melt, by numerically completing the convergent integral, we 
recovered the result reported in 
\cite{Wittmer_07a}: $\delta R^2 / R_0^2 = - 1.40/\sqrt{\bar N}$.

\bibliographystyle{ref/achemso}
\bibliography{ref/theory,ref/crossover,ref/utilities,ref/simulation}

\end{document}